\documentclass[twocolumn]{aa}  

\usepackage{graphicx}
\usepackage{txfonts}
\usepackage{lipsum}
\usepackage{subcaption}         
\usepackage{lscape}             
\usepackage{placeins}           
\usepackage[english]{babel}

\usepackage{xcolor}
\usepackage{dcolumn}
\usepackage{bm}
\usepackage{soul}
\usepackage{ulem} 

\newcommand{\dd}[0]{\ensuremath{\mathrm{d}}}
\newcommand{\ee}[0]{\ensuremath{\mathrm{e}}}
\newcommand*{\ii}{\imath}
\newcommand{\vct}[1]{\ensuremath{\bm{#1}}}%
\newcommand{\uvct}[1]{\ensuremath{\hat{\bm #1}}}

\usepackage{amsmath}
\usepackage[colorlinks=true, allcolors=blue]{hyperref}
\usepackage{listings}

\begin{document}
\date{Preprint number: LAPTH-024/26}
\title{Polarized 3D Synthetic Turbulence}
\subtitle{I. Magnetic Field Line Random Walk}

\author{Matthieu Bouchet\inst{1}\fnmsep\thanks{matthieu.bouchet@lapth.cnrs.fr}
        \and Yoann G\'enolini\inst{1}\fnmsep\thanks{yoann.genolini@lapth.cnrs.fr}
        \and Silvio S. Cerri\inst{2}\fnmsep\thanks{silvio.cerri@oca.eu}
        \and Alexandre Marcowith\inst{3}\fnmsep\thanks{alexandre.marcowith@umontpellier.fr}
        \and Philipp Mertsch\inst{4}\fnmsep\thanks{pmertsch@physik.rwth-aachen.de}
        }
  \institute{LAPTh, CNRS, USMB, F-74940 Annecy, France
   \and Université Côte d’Azur, Observatoire de la Côte d’Azur, CNRS,
Laboratoire Lagrange, Bd de l’Observatoire, CS 34229, 06304 Nice cedex 4, France \and LUPM, CNRS/Universit\'e de Montpellier, Pl. Eugène Bataillon, 34095 Montpellier, France \and Institute for Theoretical Particle Physics and Cosmology (TTK), RWTH Aachen University, 52056 Aachen, Germany}

  \abstract
   {The behavior of magnetic field lines in a turbulent plasma is a key property of the medium, with important consequences for plasma dynamics and charged-particle transport. }
   {We study the diffusion properties of magnetic field lines in synthetic turbulence featuring different polarization configurations for the magnetic perturbations, as prescribed by the existing magnetohydrodynamic modes (namely, Alfv\'enic and magnetosonic). These turbulent field realizations are then compared with the isotropic (or, random) polarization case, which is the one typically adopted in the literature.} 
   {We construct \textit{polarized} synthetic turbulence simulations and study the properties of field lines through the running diffusion coefficient.}
   {Our key findings can be summarized as follow: (i) field line wandering is strongly dependent on polarization configurations, (ii) we unveil that the sub-diffusive phase of field line is highly dependent on the polarization and is well reproduced by theoretical predictions based on Corrsin's hypothesis in the low turbulence level regime, (iii) in particular the scaling of the asymptotic diffusion coefficient in magnetosonic-like polarization is $(\delta B/B)^4$ at odd with the $(\delta B/B)^2$ scaling found in the quasi-linear regime for random polarization, (iv) interestingly we note that the subdiffusive phase of field line transport in the magnetosonic-like polarization follows closely the one observed in recent high resolution MHD turbulence simulations, we end giving a word of caution when FL transport is investigated in such simulations.}{}


\maketitle
\nolinenumbers 


\section{Introduction }

Magnetic field line random walk (FLRW) is an essential process involved in the transport of cosmic rays (CRs) or of any other charged particles such as, for instance, dust or conduction electrons in the interstellar or intergalactic medium. In general, CRs can jump from one field line (FL) to another due to scattering on background fluctuations, or simply due to the finite size of their gyro-motion. However, when their Larmor radius is sufficiently small compared to the turbulence coherence length, particles propagate along a magnetic-FL for a significant amount of time between two scattering events \citep{1995A&A...302L..21D}. Hence, how the turbulent motions of the background plasma shape and transport the magnetic FLs strongly condition the propagation of CRs \citep{1968PhRvL..21...44J, 1973ApJ...183.1029J}; see recent reviews by \citet{2020Ap&SS.365..135M, 2024AdSpR..73.1073S}, and references therein. Thus, as magnetic FLs disperse due to the chaotic nature of the turbulent eddies they intrinsically induce a perpendicular transport of CRs \citep{1966ApJ...146..480J,1978PhRvL..40...38R, 2025ApJ...992...10K, 2025ApJ...992...11K}.
 Such chaotic motions require at least three spatial dimensions (3D) to develop. In fact, 1D turbulence does not lead to perpendicular diffusion as particles stay confined in a flux tube and 2D turbulence does not have chaotic motions \citep{1998ApJ...509..238J, 2001PhRvD..65b3002C}. For instance, in pure 2D turbulence, magnetic FLs are trapped on equipotential flux surfaces, hence keeping memory of initial conditions for long distances \citep{1998ApJ...509..238J}; then adding a slab component allows to restore diffusion, but only if the amplitude of the slab component is strong enough, otherwise subdiffusion emerges instead \citep{2011ApJ...741...16G}. 
 
 The magnetic FL transport can be characterized by different means. First, as for particles, it is possible to construct an {\it absolute} diffusion coefficient based on the averaging trajectories of a given set of FLs and derive a FL running diffusion coefficient as function of the path length $s$ along the FL itself \citep{2024AdSpR..73.1073S}. This aspect of the FL transport is sensitive to the large scales involved in the turbulence. At the same time, it is also relevant to consider {\it relative} FL diffusion, i.e., how {\it two FLs} are getting separated as function of $s$. This effect controls the spread of the particles in the perpendicular direction \citep{2004ApJ...614..420R, Bouchet:2025tiv} but has also a tight relationship to the dynamics of magnetized turbulence. It can provide a framework to account for magnetic reconnection rate in magnetohydrodynamic (MHD) turbulent flows \citep{2011ApJ...743...51E} or in the process of magnetic dynamo  \citep{PhysRevE.83.056405}. The latter process is sensitive to the small scales in the turbulent cascade, an interesting aspect while considering the difference between turbulence populated by Alfvénic or magnetosonic fluctuations \citep{2002PhRvL..88x5001C}. In both approaches, i.e., either based on absolute or relative FLs diffusion, the transport coefficients do evidently depend on the properties of the turbulent spectrum (e.g., composition, anisotropy of the cascade with respect to a mean magnetic-field direction, etc). Several models have been tested analytically and numerically so far: slab and 2D turbulence, or a mixture of them as mentioned above \citep{2008JPlPh..74..657K, 2010Ap&SS.330..279S}, isotropic turbulence \citep{2001PhRvD..65b3002C, 2009JPlPh..75..183K, 2015ApJ...798...59S} including the effect of a mean field \citep{2016ApJS..225...20S}, and anisotropic turbulence \citep{2010MNRAS.406..634W}. 
However, MHD turbulence, a regime relevant for the description of CR transport, includes several specific effects like, scale-dependent anisotropy due to the very nature of Alfvénic wavepackets interaction or intermittency structures \citep{2022JPlPh..88e1501S}; all such effects are only marginally retained in the above-mentioned models \citep{2013MNRAS.431.1923S}. Interactions of Alfvén modes, whose magnetic perturbations are transverse to the background field, are likely to be the main process controlling field line random walk in strong MHD turbulence \citep{2008ApJ...673..942Y}. In the weak turbulence regime, however, MHD turbulence is also thought to be a mixture of Alfvén and magnetosonic perturbations, the latter being likely important in the modeling of parallel and perpendicular CR transport \citep{2022MNRAS.512.2111H}. Contrary to Alfv\'enic fluctuations, magnetosonic modes do have magnetic perturbation components along the background magnetic field $\vct{B}_0$, and so it is essential to retain this component to investigate FL and CR transport \citep{2009PhRvE..80f6408S}.\\

In this work, to seek a closer connection between synthetic and MHD turbulence, we do a step forward in FL transport studies by considering synthetic realizations of the turbulent magnetic field that respect the polarization associated to Alfv\'en or magnetosonic waves. This formally affects the turbulence through the turbulent correlation tensor
\begin{align}
P_{jl}(\vct{k}) \equiv 
{1 \over (2 \pi)^3} 
\int_{-\infty}^{\infty} \dd^3 \vct{\xi}
\, \ee^{\ii \vct{k} \cdot \vct{\xi} } 
\langle \delta B_j(\vct{r}) \delta B_{l}(\vct{r} - \vct{\xi}\!) \rangle \,, 
\end{align} 
i.e. the Fourier transform of the two-point correlation function, which becomes anisotropic. We produce high-resolution magnetostatic synthetic turbulence which we use to study FL transport along and across a global mean field $\vct{B}_0$. In a companion paper (Paper II), we study particle transport in the same turbulent fields, highlighting its connection to FL transport. In this paper, the results on the running FL diffusion coefficients are compared with semi-analytical calculations based on the formalism developed in \citet{2015ApJ...798...59S,2016ApJS..225...20S}, where the arc-length $s$ is replaced with a magnetic-field dependent value $\tau$ such that $\dd\tau=\dd s/|\boldsymbol{B}|$. 
We do so for various turbulence levels characterized by their Kubo number $K_\sharp = \epsilon\, {\ell_\parallel \over \ell_\perp}$, where $\epsilon=\delta B/B_0$ is the amplitude of perturbed magnetic field ($\delta B^2 \equiv \langle |\delta\vct{B}|^2 \rangle$) with respect to background field $B_0$ and $\ell_{\parallel, \perp}$ stems for the parallel or perpendicular coherence lengths of the turbulent fluctuations \citep{2009ASSL..362.....S}. However, note that in general, there is no unique diffusion coefficient value associated with a given Kubo number in the case of anisotropic turbulence \citep[see e.g.][]{2001PhRvE..63f6405P}. Hereafter, we only consider isotropic turbulence where $\ell_\parallel=\ell_\perp$.
Thus, we introduce the parameter 
\begin{equation}
\eta =\frac{\delta B^2}{\delta B^2+ B_0^2} = \frac{\epsilon^2}{1+\epsilon^2}\;, \label{eq:def_eta}
\end{equation} 
that will be used to label our results. In our case, the $\eta$ parameter is directly related to $K_\sharp= \epsilon$. We explore both small perturbation regimes with $K_\sharp \ll 1$, customarily explained with quasi-linear theory, and large perturbation regimes with $K_\sharp \ge 1$, which require non-linear theories~\citep{2024AdSpR..73.1073S}. 

In this study we find that, regardless of the turbulence level, the polarization of the turbulent fluctuations greatly affects the diffusion of FLs as a function of their arc-length $s$. In particular, in turbulence with pure magnetosonic-like fluctuations we find an enhanced sub-diffusive regime, i.e. a transient during which the mean squared displacement orthogonal to the guide field $\vct{B}_0$ behaves as:
\begin{equation}
\langle\Delta x \rangle^2 \propto s^\alpha\;,
\label{eq:diff_subdiff}
\end{equation}
with $\alpha<1$, which in turn suppresses the converged value of the diffusion coefficient by orders of magnitudes. This could have important consequences for the particle transport in such conditions. We study this behavior numerically and recover this finding with analytical arguments. \\

The paper is organized as follow: in Section~\ref{sec:II} we review the theories of magnetic FL transport in weak and strong turbulence, introducing definitions of the running diffusion coefficients; in Section~\ref{sec:III} we describe our numerical setup, the design of harmonic synthetic turbulence with polarized modes and magnetic FL reconstruction; in Section~\ref{sec:IV} we discuss our results, with a special Section~\ref{sec:V} on intriguing similarities with recent MHD results, and conclude Section~\ref{sec:VI}. The paper is supplemented by appendices to highlight properties of the different polarized turbulence considered in the paper: Appendix~\ref{app:A} describes the polarization tensors, Appendix~\ref{app:B} presents maps of magnetic fluctuations, Appendix~\ref{app:C} presents the statistics of magnetic field curvatures, Appendix~\ref{app:D} investigates the differences compares between $s$ and $\tau$ parameterization, Appendix~\ref{app:E} tests the impact of $\delta B_\parallel$ in the FL transverse diffusion, and finally Appendix~\ref{app:F} which gives more details on a specific calculation.

\section{Theory of magnetic field line transport\label{sec:II}}

\subsection{Field line transport formalism}

We define a magnetic field line as the curve $\vct{r}(s)$ that solves the set of coupled differential equations, called the field line equations 
\begin{equation}
  \frac{\dd \vct{r}(s)}{\dd s} = \frac{\vct{B}(\vct{r}(s))}{|\vct{B}(\vct{r}(s))|} \equiv \vct{b}(\vct{r}(s)) \, , 
  \label{eqn:field_line_equations-1} 
\end{equation}
with the initial condition $\vct{r}(0) = \vct{r}_0 = (x_0, y_0, z_0)^T$. 
Here, $s$ denotes again the arc length along the field line. In the following, we assume for the magnetic field $\vct{B}$ to consist of a deterministic component $\vct{B}_0 = B_0 \vct{z}$ and a zero-mean stochastic component $\delta \vct{B}$ such that $\langle \delta\vct{B} \rangle = \vct{0}$ (so that $\langle \vct{B} \rangle = \vct{B}_0$) and $\delta B\equiv \sqrt{\langle |\delta\vct{B}|^2 \rangle}$; here $\langle\dots\rangle$ denotes an ensemble average (which is replaced by a spatial average, in this case). 

For a random field $\vct{B}(\vct{r})$, the field line equation~\eqref{eqn:field_line_equations-1} is a stochastic differential equation and its solutions, the field lines $\vct{r}(s)$ themselves become random fields.
It is interesting to study the statistical properties of the field lines, for instance their statistical moments. 
Given the fact that the stochastic part of the magnetic field has zero mean, $\langle \delta \vct{B} \rangle = \vct{0}$, the mean field line position in the perpendicular directions $x$ and $y$ is always 0. 
However, the mean-square displacement $\left( \langle (x - x_0)^2 \rangle + \langle (y - y_0)^2 \rangle \right)$ is in general {\it not} zero. 
Under certain conditions and at large distances from the initial point $\vct{r}_0$, the mean-square displacement grows linearly with arc-length $s$: the hallmark of a diffusion process. 
Something similar can be found for the displacement in the parallel direction even though here the mean displacement does not necessarily vanish due to the mean field $B_0$. 
All of this motivates defining both the perpendicular and parallel running FL diffusion coefficients through the derivatives of the mean-square displacements. Using equation \eqref{eqn:field_line_equations-1} we find
\begin{align}
    {\cal D}_{\perp} &= \frac{1}{4}\frac{d}{ds}\left(\langle(x(s)-x_0)^2\rangle + \langle(y(s)-y_0)^2\rangle\right) \label{eq:def_running_s}\\ &= \frac{1}{2}\left(\langle b_x(\vct{r}(s))(x(s)-x_0) \rangle + \langle b_y(\vct{r}(s))(y(s)-y_0) \rangle\right) \label{eq:perp_rdc_s} \ , \\
    {\cal D}_{\parallel} &=\frac{1}{2} \frac{d}{ds}\langle (z(s)-\langle z\rangle)^2\rangle =\langle b_z(\vct{r}(s))(z(s)-\langle z\rangle)\rangle \ .
\end{align}

If the motion of field-lines is indeed diffusive, $\mathcal{D}_{\perp}$ and $\mathcal{D}_{\parallel}$ converge to the asymptotic FL diffusion coefficients ${ \mathcal K }_{\perp}$ and ${ \mathcal K }_{\parallel}$, respectively, 
\begin{align}
{ \mathcal K }_\perp &=  \lim_{s \to \infty } {\cal D}_\perp \ , \\
{ \mathcal K }_\parallel &=  \lim_{s \to \infty } {\cal D}_\parallel \ .
\end{align}

The fact that the stochastic field appears not just in the numerator of the right-hand-side of equation \eqref{eqn:field_line_equations-1}, but also in the denominator makes the analytical manipulation more involved. It has therefore proven beneficial~\citep{2015ApJ...798...59S} to replace the arc-length $s$ with $\tau$ defined by
\begin{equation}
\dd \tau = \dd s/|\vct{B}| \, , \label{eqn:def_tau}
\end{equation}
such that the field-line equation simplifies to 
\begin{equation}
  \frac{\dd \vct{r}(\tau)}{\dd \tau} = \vct{B}(\vct{r}(\tau)) \, . 
  \label{eqn:field_line_equations-2} 
\end{equation}
The field line is now a function of $\tau = \int \dd s / B(\vct{r}(s))$. 
We redefine the running diffusion coefficients accordingly,
\begin{align}
    {D}_{\perp} &= \frac{1}{4}\frac{d}{d\tau}\left(\langle(x(\tau)-x_0)^2\rangle + \langle(y(\tau)-y_0)^2\rangle\right)\\ 
    &= \frac{1}{2} \left( \langle B_x(\vct{r}(\tau))(x(\tau)-x_0) \rangle + \langle B_y(\vct{r}(\tau))(y(\tau)-y_0) \rangle \right) \ , \label{eq:perp_rdc_tau} \\
    {D}_{\parallel} &= \frac{1}{2}\frac{d}{d\tau}\langle (z-\langle z\rangle)^2\rangle = \langle B_z(\vct{r}(\tau)) (z-\langle z \rangle )\rangle \ .
\end{align}
with asymptotic values 
\begin{align}
{K}_\perp &=  \lim_{\tau \to \infty } D_\perp \ , \\
{K}_\parallel &= \lim_{\tau \to \infty } D_\parallel \ .
\label{eq:limits}
\end{align}
The relation  between K and ${ \mathcal K }$ is found to be $K \approx  B~ {\mathcal K}$  in the limit of small turbulence levels. A more general relation can be found in the random ballistic diffusion (RBD) closure by \citet{2015ApJ...798...59S,2016ApJS..225...20S} (see Sect.~\ref{sec:closure}). Note that the mean-field has $z$-coordinate $\langle z \rangle = B_0 \tau$. 

\subsection{Quasi-linear theory of field line transport}\label{sec:qlt}
For low turbulence level, $\delta {B} \ll {B}_0$, one can approximate the arc-length $s$ with the distance in the $z$-direction. We can further ignore the $z$-component of the stochastic field with respect to the deterministic part $B_0$. 
The field line equation \eqref{eqn:field_line_equations-1} thus simplifies significantly, 
\begin{equation}
  \frac{\dd x}{\dd z} = \frac{\delta B_x(z)}{B_0} \, , \quad \frac{\dd y}{\dd z} = \frac{\delta B_y(z)}{B_0} \, . 
\end{equation}

Here we follow \citet{2016ApJS..225...20S} in adopting a Taylor-Green-Kubo (TGK) approach~\citep{taylor1922,1954JChPh..22..398G,1957JPSJ...12..570K}. Integrating the field-line equation in the $x$-direction, for instance, we can compute the mean-square displacement $\sigma_x^2=\langle (\Delta x)^2 \rangle$ as 
%
\begin{align}
  \sigma_x^2 &= \frac{1}{B_0^2} \int_0^z \dd z' \int_0^z \dd z'' \langle \delta B_x(z') \delta B^*_x(z'')\rangle \\
  &= \frac{1}{B_0^2} \int \dd\vct{k}' \int \dd\vct{k}'' \int_0^z \dd z' \int_0^z \dd z''\langle \tilde{{\cal C}}(z',z'',\vct{k}',\vct{k}'') \rangle \, , \label{eqn:sigma_x_squared_QLT-1}
\end{align}
where $\tilde{{\cal C}}(z',z'',\vct{k}',\vct{k}'')= \delta \tilde{B}_x(\vct{k}') \delta \tilde{B}_x(\vct{k}'') \ee^{\ii (\vct{k}' \cdot \vct{r}(z') - \vct{k}'' \cdot \vct{r}(z''))}$ contains the Fourier-transform pair for the turbulent magnetic field, $\delta \vct{B}(\vct{r}) = \int \dd\vct{k}  \, \ee^{\ii \vct{k} \cdot \vct{r}} \delta \tilde{\vct{B}}(\vct{k})$ and $\delta \tilde{\vct{B}}(\vct{k}) = \frac{1}{(2\pi)^3} \int \dd\vct{r} \, \ee^{-\ii \vct{k} \cdot \vct{r}} \delta \vct{B}(\vct{r})$.

Adopting Corrsin's independence hypothesis~\citep{1959AdGeo...6..161C}, considering the turbulence to be homogeneous and assuming for the characteristic functions in the parallel and perpendicular directions to be independent, equation \eqref{eqn:sigma_x_squared_QLT-1} simplifies to \citep[for details, see sections 1.3.2 and 2.4.4 of][]{2009ASSL..362.....S}
\begin{align}
\sigma_x^2 &\simeq \frac{2}{B_0^2} \int \dd\vct{k} \int_0^z \dd z' (z-z') \, \langle \delta \tilde{B}_x(\vct{k}) \delta \tilde{B}^*_x(\vct{k}) \rangle \langle \ee^{\ii \vct{k} \cdot \vct{\Delta r}} \rangle \\
&\simeq \frac{2}{B_0^2} \int \dd\vct{k} \int_0^z \dd z' (z-z')\,\langle \delta \tilde{B}_x(\vct{k}) \delta \tilde{B}^*_x(\vct{k}) \rangle \cos \left(k_{\parallel} z' \right) \nonumber \\ 
& \qquad \times \langle \ee^{\ii \vct{k}_{\perp} \cdot \vct{\Delta r}_{\perp}} \rangle. \label{eqn:ansatz_ODE_method}
\end{align}

Differentiating the above expression twice with respect to $z$ results in the set of coupled differential equations for $\sigma_x^2$ and $D_{\perp}$,
\begin{align}
{\dd \,\sigma_x^2\over \dd z} &= 2\, {\cal D}_{\perp} \ ,\label{eq:sigx2_Dperp_s} \\
{\dd \,{\cal D}_{\perp}\over \dd z}   &= {1 \over B_0^2} \int \dd\vct{k}\, P_{xx}(\vct{k})\,{\cos(k_\parallel z) \exp(-\sigma_x^2 k_\perp^2/2)} \ .\label{eq:Dperp_int_s}
\end{align}

Here, we have identified $\langle \delta \tilde{B}_x(\vct{k}) \delta \tilde{B}^*_x(\vct{k}) \rangle$ as the $xx$-component of the turbulence tensor $P_{xx}$ and employed the characteristic function for a Gaussian distribution with variance $\sigma_x^2$, 
\begin{equation}
\langle \ee^{\ii \vct{k}_{\perp} \cdot \vct{\Delta r}_{\perp}} \rangle = \exp(-\sigma_x^2 k_\perp^2/2) \, . 
\end{equation}
This exponential term is typically neglected in the so-called quasi-linear approximation~\citep{2021PhPl...28l0501S}.

\subsection{Strong turbulence}\label{S:STU}
In the strong turbulence regime, i.e. when $\delta B \gtrsim B_0$, adopting the $\tau$ formalism introduced in Eq.~\eqref{eqn:field_line_equations-2} leads to more accurate semi-analytical predictions for the FL transverse transport~\citep{2015ApJ...798...59S}. In particular this formalism allows to naturally include magnetic perturbations along the background magnetic field, namely $\delta B_z \neq 0$ \citep{2009PhRvE..80f6408S}. In that case, a calculation similar to the one conducted in section~\ref{sec:qlt} --- i.e., still using the Corrsin's hypothesis, but now implementing the variable $\tau$ --- leads to a system of four coupled ODEs to include transport along the background magnetic field:
\begin{align}
{\dd \,\sigma_x^2\over \dd \tau} &= 2\, { D}_{\perp} \ , \label{eq:sigx2_Dperp_tau} \\
{\dd \,{ D}_{\perp}\over \dd \tau}   &=  \int \dd\vct{k} P_{xx}(\vct{k}){\cos(B_0 k_\parallel \tau) \exp\left(-{\sigma_x^2 k_\perp^2 \over 2}-{\sigma_z^2 k_\parallel^2 \over 2}\right)} \ , \label{eq:Dperp_int_tau} \\
{\dd \,\sigma_z^2\over \dd \tau} &= 2\, { D}_{\parallel} \ ,\label{eq:sigz2_Dpara_tau} \\
{\dd \,{ D}_{\parallel}\over \dd \tau}   &= \int \dd\vct{k} P_{zz}(\vct{k}){\cos(B_0 k_\parallel \tau) \exp\left(-{\sigma_x^2 k_\perp^2 \over 2}-{\sigma_z^2 k_\parallel^2 \over 2}\right)} \ .\label{eq:Dpara_int_tau}
\end{align}
where $\sigma_x^2=\langle (\Delta x)^2 \rangle$ and $\sigma_z^2=\langle (\Delta z)^2 \rangle$. 

\subsection{Closure methods \label{sec:closure}}
In general, determining FL transport from the systems of equations in $z$ (Eqs.~\eqref{eq:sigx2_Dperp_s}-\eqref{eq:Dperp_int_s}), or in $\tau$ (Eqs.~\eqref{eq:sigx2_Dperp_tau}-\eqref{eq:Dpara_int_tau}) can be done by solving the corresponding ODE system numerically. Such an approach enables obtaining the detailed shape of the running FL diffusion coefficients and leads to the most accurate results, provided the underlying assumptions are correct. These are the reasons why we mostly discuss the results from this method in Sect.~\ref{sec:IV}. Nonetheless, several closure relations have been proposed in the literature to provide an analytical description of the asymptotic values of the FL diffusion coefficient $\cal K$ or $K$ in our notations.\\ 

In the \textit{diffusive decorrelation} (DD) closure introduced by \citet{1995PhRvL..75.2136M}, one assumes a diffusive spreading of the FLs at all $z$ or $\tau$, such that $\sigma_x^2= 2{\cal K_\perp} z$ or $2 K_\perp \tau$, respectively. Using this prescription in the two remaining Eqs.~\eqref{eq:Dperp_int_tau}-\eqref{eq:Dpara_int_tau}, leads to the following system~\citep{2016ApJS..225...20S}:
\begin{eqnarray} {K}_{\perp} & =  \displaystyle \int \dd{\vct{k}} {P}_{{xx}}({\vct{k}})\displaystyle \frac{{K}_{\perp}{k}_{\perp}^{2}+{K}_{\parallel}{k}_{\parallel}^{2}}{{({K}_{\perp}{k}_{\perp}^{2}+{K}_{\parallel}{k}_{\parallel}^{2})}^{2}+{B}_{0}^{2}{k}_{\parallel}^{2}} \label{eq:Kperp_DD} \ ,\\
{K}_{\parallel} & =  \displaystyle \int \dd{\vct{k}} {P}_{{zz}}({\vct{k}})\displaystyle \frac{{K}_{\perp}{k}_{\perp}^{2}+{K}_{\parallel}{k}_{\parallel}^{2}}{{({K}_{\perp}{k}_{\perp}^{2}+{K}_{\parallel}{k}_{\parallel}^{2})}^{2}+{B}_{0}^{2}{k}_{\parallel}^{2}} \ . \label{eq:Kpar_DD}
\end{eqnarray}

In the \textit{random ballistic decorrelation} (RBD) closure introduced by \citet{2011ApJ...741...16G}, one  assumes a ballistic spreading of the FLs over the distance scales in $z$ or $\tau$ relevant to decorrelation of the random walk. In that case $\sigma_x^2= \langle \delta B_x^2 \rangle z^2 /B_0^2 $ or $\langle \delta B_x^2 \rangle \tau^2$, respectively. Using this prescription in the two remaining Eqs.~\eqref{eq:Dperp_int_tau}-\eqref{eq:Dpara_int_tau}, lead to the following independent equations~\citep{2016ApJS..225...20S}:

\begin{eqnarray}
{K}_{\perp}=\displaystyle \frac{1}{2}\int \dd{\vct{k}}\,\sqrt{\displaystyle \frac{\pi }{{\nu }_{2}}}{P}_{{xx}}(\vct{k}){e}^{-\tfrac{{B}_{0}^{2}{k}_{\parallel}^{2}}{4{\nu }_{2}}} \ ,\label{eq:Kperp_RBD}\\
{K}_{\parallel}=\displaystyle \frac{1}{2}\int \dd{\vct{k}}\, \sqrt{\displaystyle \frac{\pi }{{\nu }_{2}}}{P}_{{zz}}(\vct{k}){e}^{-\tfrac{{B}_{0}^{2}{k}_{\parallel}^{2}}{4{\nu }_{2}}} \ ,\label{eq:Kpar_RBD}
    \end{eqnarray}
with ${\nu }_{2}\equiv \tfrac{1}{2}(\langle {\delta B}_{x}^{2}\rangle {k}_{\perp}^{2}+\langle {\delta B}_{z}^{2}\rangle {k}_{\parallel}^{2})$.\\

These closed-form solutions (i.e., DD and RBD) for the asymptotic diffusion coefficients are particularly powerful in predicting the quasi-linear scalings and normalization for low turbulence levels (small $\eta$) regime. They will be used for this purpose in Sect.~\ref{sec:DD_RBD}. Note that in this regime these two prescriptions lead to the same analytical value~\citep{2016ApJS..225...20S}.  

In this part, we have not detailed the case for the $z$ variable, but the systems of equations obtained in the RBD or DD closures are similar. Interestingly, a relation between $K$ and ${ \mathcal K }$ can be found within the RDB closure~\citep{2016ApJS..225...20S}: ${K}_{i}(\tau )/{{ \mathcal K }}_{i}(z)\approx \langle {\delta B}_{i}^{2}/| B| \rangle /\langle {\delta B}_{i}^{2}/{B}^{2}\rangle $. 

\section{Polarized Synthetic Turbulence\label{sec:III}}

\subsection{Formalism}
To study magnetic FL transport we generate synthetic magnetostatic turbulent fluctuations as a sum of plane waves with random phases~\citep{1994ApJ...430L.137G}. Motivated by the different nature of the various MHD modes we explicitly distinguish between their various polarization for the magnetic perturbations.
Each wave is characterized by a wave vector $\vct{k}=k\uvct{k}$ and perturbation amplitude $\vct{\delta \tilde{B}(\vct{k})}=\delta \tilde{B}\uvct{\xi}$ (hatted symbols correspond to unit vectors). Given the background magnetic field $\vct{B_0}=B_0\uvct{z}$, we can define two transverse directions to the wave vector, $\uvct{y}'= {\uvct{z} \times \uvct{k}}/{|\uvct{z} \times \uvct{k}|} $ and $\uvct{y}'\times \uvct{k}$ on which the perturbation can be projected such that: $\vct{\delta \tilde{B}}=\delta \tilde{B}(\cos \alpha \,\uvct{y}'\times \uvct{k} + \sin \alpha\, \uvct{y}')$. In the following we will consider turbulence with either random values for $\alpha$ (ISO case, as \textit{isotropic variance}), or fixed values of $\alpha$ for two other extreme cases motivated by linearized ideal-MHD wave polarization: $\alpha=\pi/2$ for Alfvèn wave polarization (ALF), and $\alpha=0$ for magnetosonic wave polarization (MAG). Note that, in the latter case, the magnetic perturbations have a component along $\vct{z}$, whereas in the former case they remain in the $(\vct{x}, \vct{y})$ plane.
Such a turbulence is characterized by the full turbulent correlation tensor $P_{jl}({\bf k})$ which can be cast into the following form for homogeneous and isotropic turbulence (see Appendix.~\ref{app:A} for details):
\begin{align}
 \langle \delta \tilde{B}_j(\vct{k}) \delta \tilde{B}^*_l(\vct{k'}) \rangle& =P_{jl}(\vct{k})\delta(\vct{k'}-\vct{k})\\ &= P(k) \langle \hat{\xi}_j \hat{\xi}_l \rangle (\uvct{k}) \delta(\vct{k'}-\vct{k})\;,\label{eq:turb_corr_tensor}
\end{align}
with $\langle \hat{\xi}_i \hat{\xi}_l \rangle$ being the polarization tensor, i.e. the ensemble average of modes' polarization vector components with a given $k$. In this case the trace of $\langle \hat{\xi}_i \hat{\xi}_l \rangle$ is simply 1, so we get the identity
\begin{equation}
\delta B^2 = \int \dd \vct{k} \,P(k)\;. \label{eq:normalization} 
\end{equation}
The polarization tensor can be parameterized as a function of $\rho$ which amounts for the contribution of the fluctuation energy imparted into the two transversal directions $\uvct{y}'= {\uvct{k} \times \uvct{z}}/{|\uvct{k} \times \uvct{z}|} $ and $\uvct{y}'\times \uvct{k}$. We denote their respective amplitudes as $\delta B^2_{\uvct{y}'}$
and $\delta B^2_{\uvct{y}'\times \uvct{k}}$. Hence, 
\begin{equation}
\rho = { \delta B^2_{\uvct{y}' } \over \delta B^2_{\uvct{y}'}+ \delta B^2_{\uvct{y}'\times \uvct{k}}}.\label{eq:def_rho}
\end{equation}
Following \citet{2016ApJS..225...20S}, the diagonal terms of the polarization tensors can be written:
\begin{align}
    \langle \hat{\xi}_x \hat{\xi}_x \rangle &=\langle \hat{\xi}_y \hat{\xi}_y \rangle = \frac{\rho}{2} +\frac{(1-\rho)}{2} \cos^2\theta \\
    \langle \hat{\xi}_z \hat{\xi}_z \rangle &= (1-\rho)(1-\cos^2\theta)
\end{align}
where $\theta$ is the angle between $\vct{k}$ and $\vct{B}_0$ and $\rho=$ 0, 0.5 and 1 correspond to MAG, ISO and ALF polarization, respectively.
From this polarization tensor we readily compute the partition of the magnetic field amplitude along the direction parallel and perpendicular to the background field $\vct{B}_0$. We find that
\begin{align}
    \delta B_\perp^2 &\equiv \langle \delta B_x^2 \rangle + \langle \delta B_y^2 \rangle = \left(\frac{1}{3}+\frac{2}{3}\rho\right)\; \delta B^2 \quad\text{and,}\nonumber\\ \delta B_\parallel^2 &\equiv\langle \delta B_z^2 \rangle= \left(\frac{2}{3}-\frac{2}{3}\rho\right)\;\delta B^2,\;\label{eq:equip}
\end{align}
Equipartition of $\delta B^2$ along the 3 coordinates is only recovered in the ISO case, for the ALF case the perturbations only project onto the $(x,y)$ plane, while for the MAG case the perturbations mostly project along the $z$-direction.

\subsection{Harmonic method}
We use the harmonic method to generate a 3D turbulent magnetic field, as originally proposed in \citep{1994ApJ...430L.137G}. Compared to grid-based methods~\citep[see, e.g.,][]{2002GeoRL..29.1048Q}, the harmonic method is better suited for this study, as it allows us to investigate field line transport over multiple correlation lengths without being limited by the grid's dynamical range. We have, however, verified consistency with the grid method wherever applicable. The local magnetic field is computed by the sum of a background field oriented along $\vct{z}$ and a perturbation $\vct{B}= B_0\uvct{z}+\vct{\delta B} $, the latter corresponding to a sum over $N$ plane waves,
\begin{align}
    \boldsymbol{\delta B}(\boldsymbol{r}) &= \sqrt{2} \sum_{n=0}^{N-1} A_n \boldsymbol{\hat{\xi}}_n \cos{(k_n \boldsymbol{\hat{k}}_n \cdot \boldsymbol{r} + \beta_n)}\;,
\end{align}
with wave vectors, 
\begin{align}
  \boldsymbol{\hat{k}}_n &= \begin{pmatrix} \sqrt{1-\chi^2_n} \cos{\phi_n} \\ \sqrt{1-\chi^2_n} \sin{\phi_n} \\ \chi_n\end{pmatrix}\;.
\end{align}
We assume the phases of the modes to be uncorrelated and thus $\beta_n$ is drawn randomly within $[0,2\pi]$. Furthermore, we assume there is no preferred direction for the wave vectors and so the angles $\chi_n=\cos \theta_n$ and $\phi_n$ are also chosen randomly from the intervals $[-1,1]$ and $[0,2\pi]$, respectively. The polarization vector, which sets the orientation of the magnetic perturbation, is a critical quantity in this study and takes the general form:
\begin{align}
   \boldsymbol{\hat{\xi}}_n &= \begin{pmatrix}
        -\sin{\alpha_n}\sin{\phi_n} + \cos{\alpha_n}\cos{\phi_n} \chi_n \\
        \sin{\alpha_n} \cos{\phi_n} + \cos{\alpha_n} \sin{\phi_n} \chi_n \\
        -\sqrt{1-\chi^2_n} \cos{\alpha_n}
    \end{pmatrix}\;.\label{eq:polarization_vector}
\end{align}
Its direction is contained in the plane orthogonal to $\vct{k}$, fixed by the choice of $(\chi_n, \phi_n)$. The direction in this plane is fixed by the angle $\alpha_n$. In previous works it was commonly drawn randomly within $[0,2\pi]$, and we refer to this case as the \textit{isotropic variance case} introduced before and abbreviated as ISO. As presented in the previous section, in this paper we are studying two other cases with fixed $\alpha_n$: the Alfvèn wave polarization (ALF) case for which $\alpha_n=\pi/2$, and the magnetosonic wave polarization (MAG) setup that corresponds to $\alpha_n=0$. The amplitude of the waves is controlled by $A_n$:
\begin{equation}
    A_n^2 = \delta B^2\; G_n\left[\sum_{n=0}^{N-1} G_n\right]^{-1}\;, 
\end{equation}
where $G_n=4\pi k^2_n \Delta k_n P(k_n)$. We consider a Kolmogorov-like power spectrum which extend from $k_{\rm min}$ to $k_{\rm max}$:
\begin{equation}
P({k}) = {\cal N} \left(\frac{k}{k_{\rm min}}\right)^{-\frac{11}{3}}\;,
\end{equation}
whose normalization ${\cal N}$ is set by the turbulence level $\delta B^2$, through Eq.~\eqref{eq:normalization}, and is equal to $\delta B^2/(6 \pi k_{\rm min}^3)$ for infinitely large $k_{\rm max}$.

Practically, the synthetic turbulent fields are generated with a dynamical range $k_{\rm max}/k_{\rm min}=256$ and a number of modes $N$ from 5000 to 10000, evenly spaced in logarithmic scale. These parameters were chosen to ensure the numerical convergence of the field line running diffusion coefficient shown in the next section. We consider different turbulent levels, where we fix $\delta B^2=(1\,\rm\mu G)^2$ and vary $B_0=10^{\nu}\rm\mu G$, with $\nu$ being an integer multiple of 0.25. Note however that our results will be presented in terms of the scale invariant quantity $\eta$ (see Eq.~\eqref{eq:def_eta}). Within these turbulent fields, we integrate Eqs. \eqref{eqn:field_line_equations-1} and \eqref{eqn:field_line_equations-2} to reconstruct field lines, using a fourth-order Runge-Kutta algorithm, with a step size chosen to be a tenth of the smallest scale of the turbulence ($l_{\rm min}=2\pi/k_{\rm max}$) to ensure accuracy. Accurately estimating the running diffusion coefficient of FLs numerically requires reconstructing a large number of FLs to compute a reliable ensemble average. This embarrassingly parallel problem is farmed out using graphics processing units (GPUs). We reconstruct about 50\,000 FLs in 10 different magnetic configurations for each turbulence level presented in the next section.  

\begin{figure}[tbh]
\includegraphics[scale=0.70]{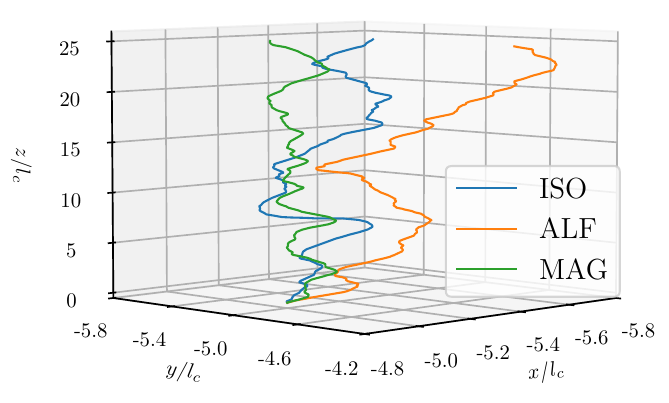}
\includegraphics[scale=0.70]{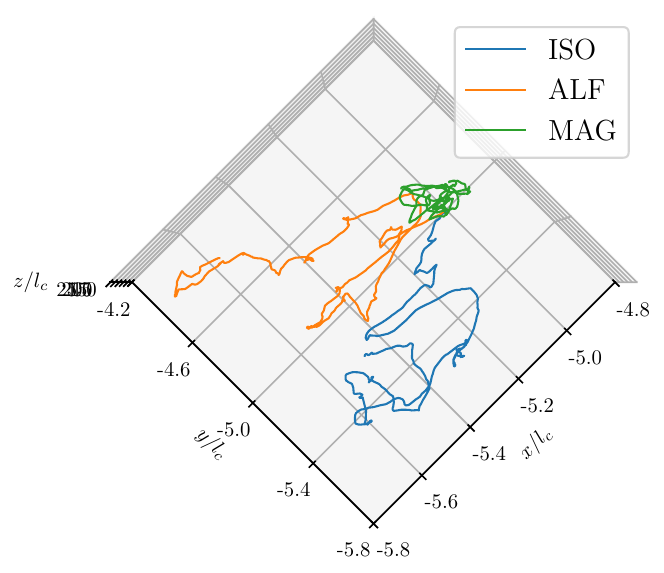}
\caption{3D integrated magnetic field lines trajectories plotted over few correlations lengths for $\eta=0.03$. The three colors correspond to the three polarization cases used in the paper: isotropic, Alfvènic and magnetosonic. Each case corresponds to an independent realization of the turbulence. Top plot: Z profiles. Bottom plot: X-Y plan projection.} 
\label{fig:field_line_wandering_3D} 
\end{figure}

\section{Results  \label{sec:IV}}
\subsection{Anisotropies in magnetic field correlations} \label{sec:aniso_lbperp}
The qualitative behavior of FL wandering in the three polarizations configurations (ISO, ALF, MAG) is drastically different. We illustrate this in Fig.~\ref{fig:field_line_wandering_3D} where we display the integration of one representative field line over a few correlation lengths $l_c$ for three different realizations of low-amplitude turbulence ($\eta=0.03$) corresponding to the three polarizations. We note that the transverse excursions of FL in the Alfvènic case are larger than the ones in the isotropic case, itself larger than the ones in the magnetosonic case. This hierarchy is definitely related with the turbulence structure: the interference of specifically polarized plane wave modes lead to different magnetic field patterns, despite the modes being distributed isotropically in wave-vector space for all polarizations. 
To complete this qualitative description, a comparison of 2D sections of the B-field are shown in App.~\ref{app:B}, and reveal different characteristic patterns for each polarization configuration.\\

\begin{figure}[tbh]
\includegraphics[scale=0.50]{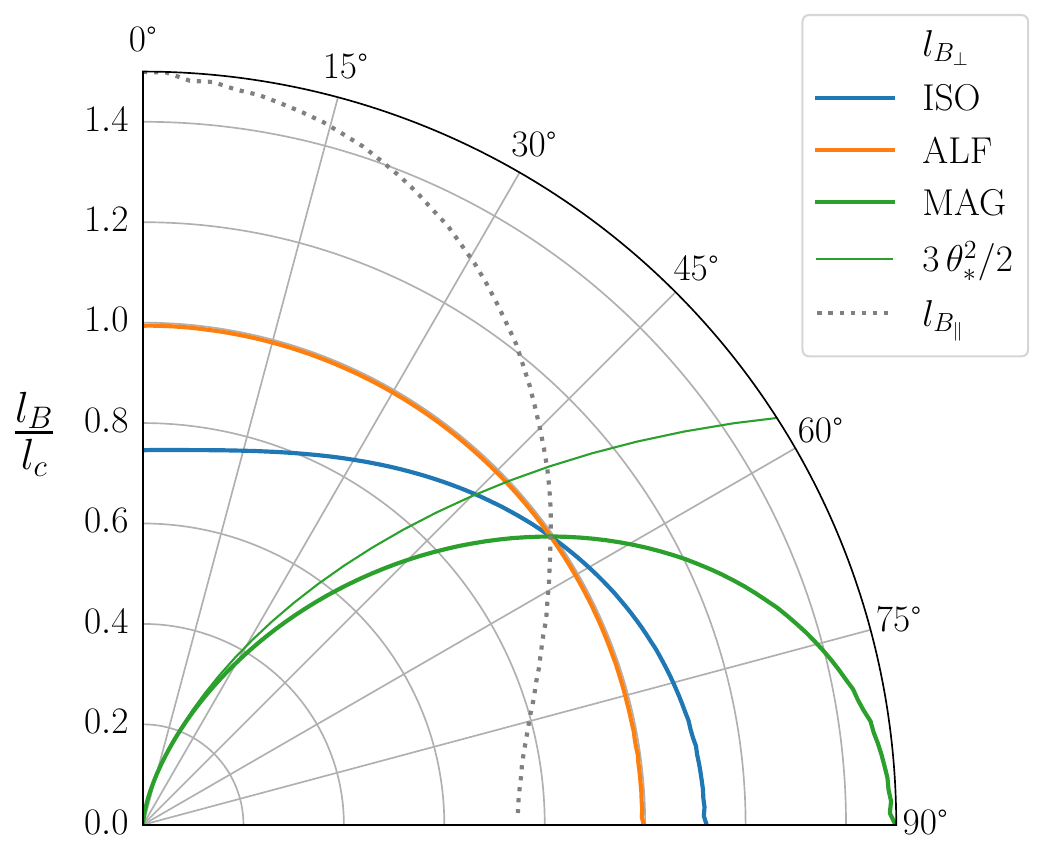}
\caption{Correlation length $l_{B_{\perp}}(\theta_*)$ (see Eq.~\eqref{eq:lperp}) and $l_{B_{\parallel}}(\theta_*)$ (see Eq.~\eqref{eq:lpara}) represented in polar coordinates for the polarization configurations discussed in the paper. Note that $l_{B_{\parallel}}$ is only relevant for the ISO and MAG cases for which it is the same.} 
\label{fig:correlation_length} 
\end{figure}

Interestingly, the correlation length of the total magnetic field fluctuations being defined as:
\begin{align}
 l_c & = \displaystyle\frac{2}{\delta B^2} \int_0^\infty \,  \dd l \, \langle \delta \vct{B}(\vct{r})  \delta \vct{B}(\vct{r} + \vct{l})\rangle \\
 &= \displaystyle\frac{2}{\delta B^2} \int_0^\infty \dd l \int \dd\vct{k}\, \left( \sum_{j=(x,y,z)} P_{jj}(\vct{k}) \right)\,e^{i\vct{k}\cdot\vct{l}}  \,\\
 &= \displaystyle\frac{4 \pi^2}{\delta B^2} \int  \dd k k  \, P(k) \,\\
 &= \displaystyle \pi \int \dd k k \, P(k) \Bigg/ \int \dd k k^2 \, P(k) \,
 \label{eq:lc}  
\end{align}
 is exactly the same for the three polarizations configurations ISO, ALF and MAG, and approximates to $l_c\approx l_{max}/5$ for the Kolmogorov turbulence spectrum considered in this paper when $l_{min}\to 0$. Note that this characteristic length is isotropic, consistent with the assumption of isotropic turbulence. However, the transverse magnetic perturbation $\delta\vct{B}_\perp$ is the one driving the transverse FL wandering, especially at small turbulence level where $|\delta\vct{B}_\parallel|\ll B_0$. We will discuss this statement more quantitatively in Sect.~\ref{sec:FL_sim} (see notably App.~\ref{app:E}).  To highlight the differences in the fluctuations $\delta \vct{B_\perp}$ in the three polarizations cases, we study the correlation length $l_{B_\perp}$ of the transverse component of the magnetic field, which is defined as: 
\begin{equation}
 l_{B_\perp}(\theta_*) = \frac{2}{\delta B_\perp^2} \int_0^\infty \langle \delta \vct{B_\perp}(\vct{r})  \delta \vct{B_\perp}(\vct{r} + \vct{l}(\theta_*))\rangle  \,  \dd l \;,\label{eq:lperp}     
\end{equation}
where the direction of $\vct{l}$ is fixed and defined by the angle $\theta_*$ measured from the $\vct{z}$-axis, corresponding to the direction of $\vct{B}_0$. This quantity serves as a proxy for the commonly used structure function, but restricted to the transverse component of the magnetic field. We compute $l_{B_\perp}$ for various orientations and report their values in polar coordinates in Fig.~\ref{fig:correlation_length}, normalized to the total correlation length $l_c$. We can see that the correlation length $l_{B_\perp}$ is isotropic and equal to $l_c$ in the ALF case, while it is anisotropic in the two other cases: within [0.75,1.15]$l_c$ in the ISO case, and within [0,1.5]$l_c$ in the MAG case. This hierarchy can already explain the differences observed in Fig.~\ref{fig:field_line_wandering_3D} for the three polarizations, and more importantly the difference in their statistical behavior that will be presented in the next section. In particular, for small turbulence level in the MAG case, FL are moving very close to the $\vct{z}$ direction (with typical $\theta_*\sim \delta B_\perp/B_0 = \delta B/(\sqrt{3} B_0)\ll 1$), for which the correlation length (and so the diffusion) is strongly suppressed. In the limit of small $\theta_*$ (i.e. small perturbation amplitudes relative to the background field), the Taylor expansion to the second order of $l_{B_\perp}$ writes:
\begin{equation}
    l_{B_{\perp}}(\theta_*)=l_{B_{\perp}}(0)+\theta_*\left.\frac{\partial l_{B_{\perp}}}{\partial\theta_*}\right|_{0}+\frac{\theta_*^2}{2}\left.\frac{\partial^2 l_{B_{\perp}}}{\partial\theta_*^2}\right|_{0}+o(\theta_*^3)
\end{equation}
In the MAG case, we can easily show that the zeroth- and first-order terms vanish, while the second-order term is nonzero and equal to $3\,l_c$. Hence we get the following scaling:
\begin{equation}
l_{B_\perp^{MAG}}(\theta_*\to 0)\approx  \frac{3}{2}\,l_c\, \theta_*^2 \approx \frac{3}{2}\,l_c\,\left( \frac{\delta B }{B_0}\right)^2\;. \label{eq:scaling_lBperp} 
\end{equation}
This function is represented with a thin green line on Fig.~\ref{fig:correlation_length} and shows that this approximation indeed works well for $\theta_* \leq 0.5 $ (in radian). For completeness, we also compute and show:
\begin{equation}
 l_{B_\parallel}(\theta_*) = \frac{2}{\delta B_\parallel^2} \int_0^\infty \langle \delta \vct{B_\parallel}(\vct{r})  \delta \vct{B_\parallel}(\vct{r} + \vct{l}(\theta_*))\rangle  \,  \dd l \;,\label{eq:lpara}     
\end{equation}
 with a dotted gray line on the same figure. Note that this quantity is exactly the same in the two cases where it is relevant (ISO, MAG). In these cases the identity $\delta B^2\; l_c= \delta B_\perp^2 \;l_{B_\perp} + \delta B_\parallel^2 \;l_{B_\parallel}$ hold for any $\theta_*$. This relation also explains why all the curves cross at the same value $l_{B_\parallel}=l_{B_\perp}=l_c$, for an angle conjectured to be $\arctan(\sqrt{2})\approx 54.73^o$.
In App.~\ref{app:C} we also characterize the curvature statistics of the three polarization cases, showing significant differences. However, inferring properties of the FLRW from this latter quantity does not seem appropriate in our case. Indeed, FL transport is mainly driven by large scale fluctuations, while FL curvature is a local property dominated by small scales (at least for small $\eta$ values); and in synthetic turbulence large and small scales are uncorrelated. Note, however, that the distribution of FL curvature has recently been shown to impact charged particle scattering in MHD turbulence~\citep[see, e.g.,][]{2023JPlPh..89e1701L,2023MNRAS.525.4985K}. 

\begin{figure*}[tbh]
\includegraphics[scale=0.326]{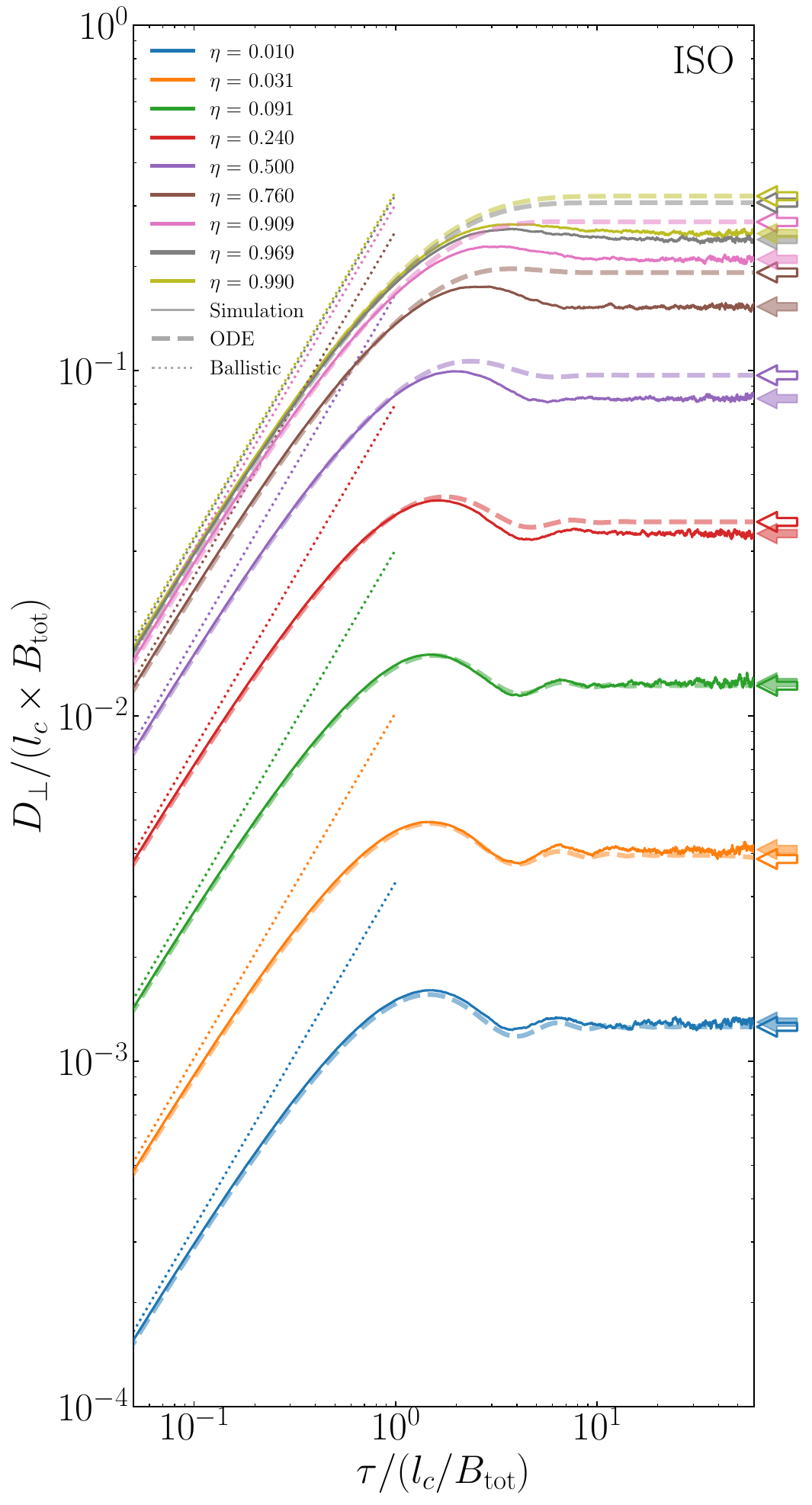}
\includegraphics[scale=0.326]{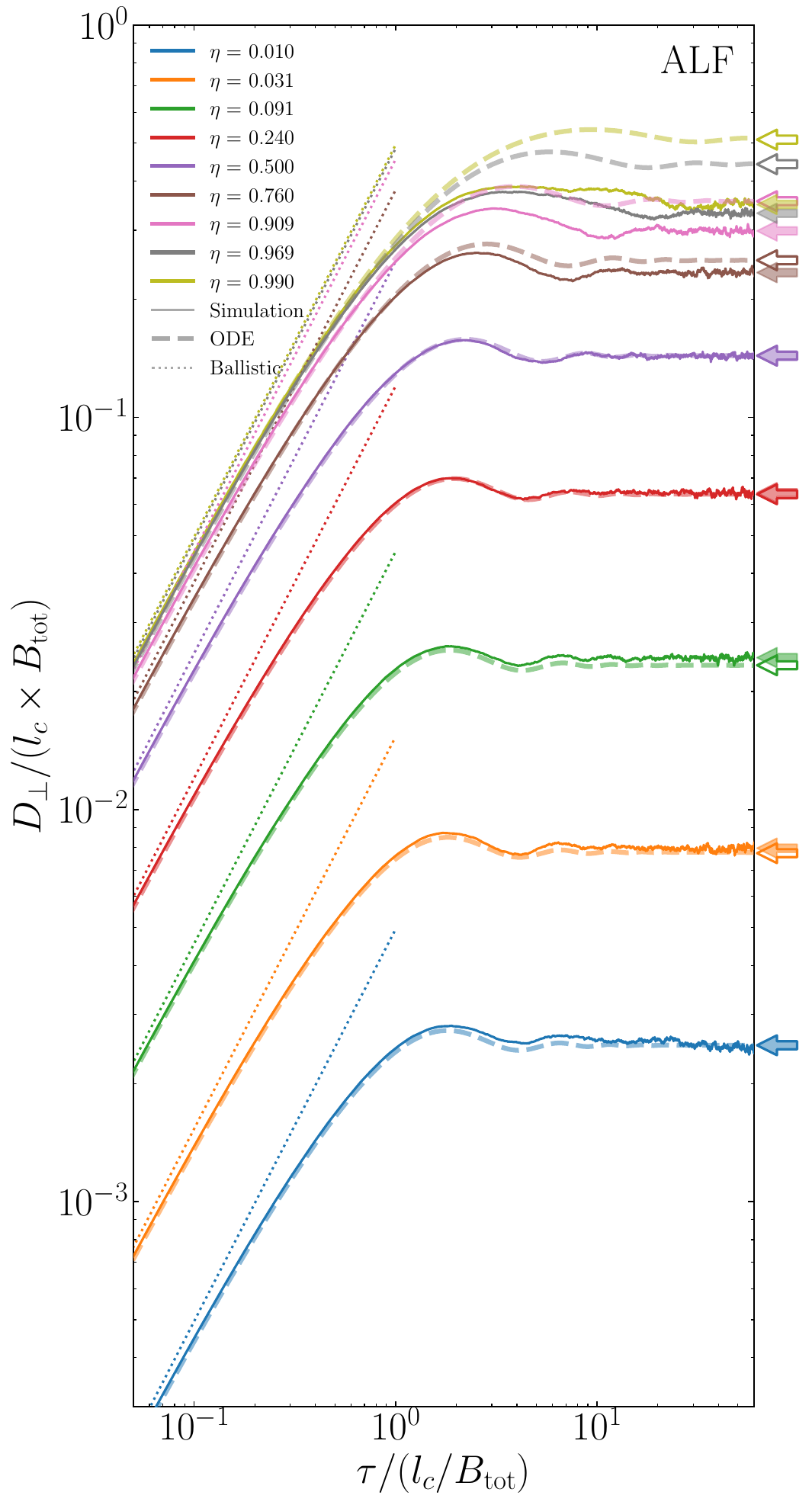}
\includegraphics[scale=0.326]{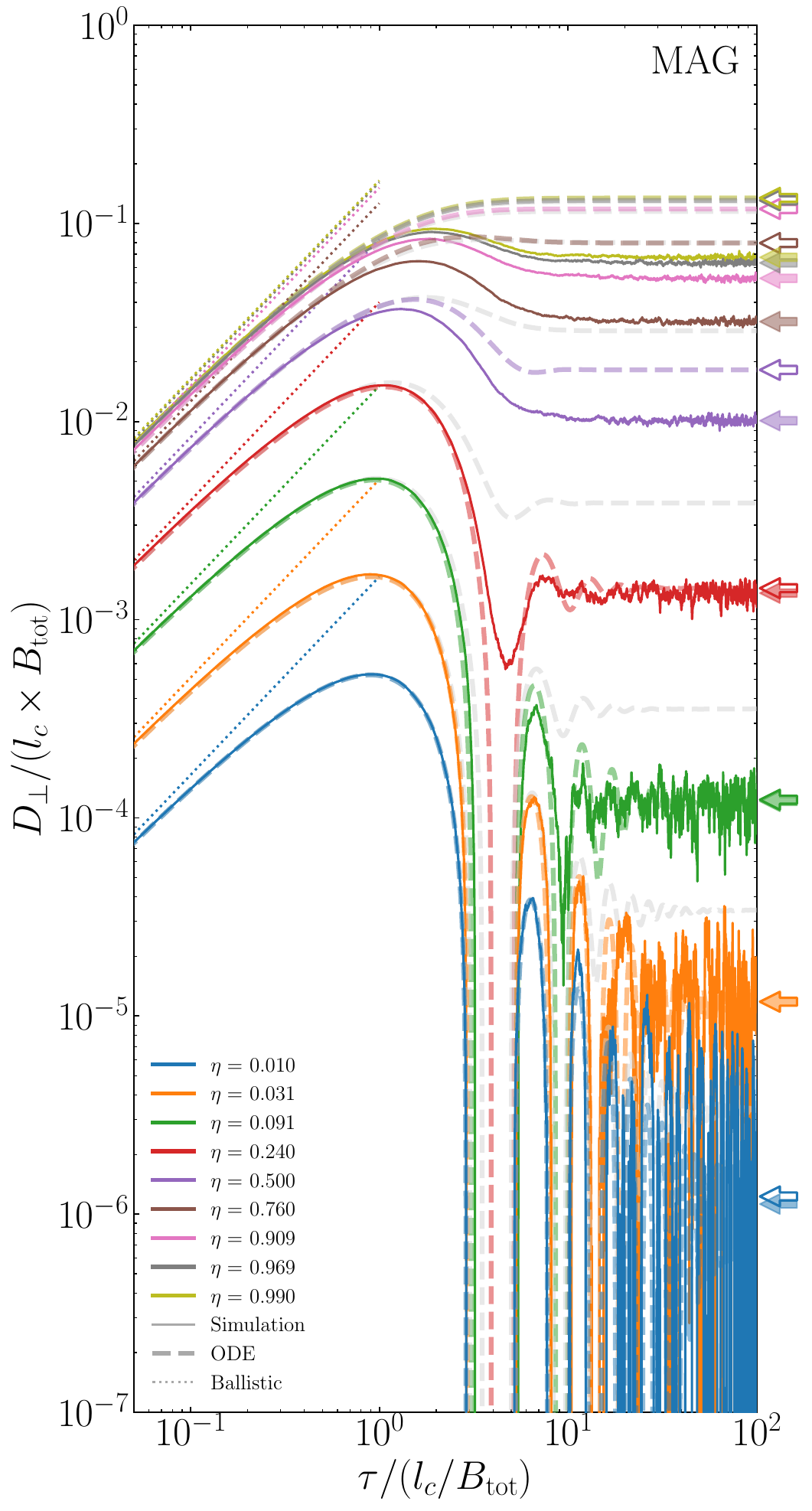}
\caption{Running FL diffusion coefficient ${ D}_{\perp}$ as a function of $\tau$ for the three polarization cases considered in the paper: ISO, ALF and MAG. The different colors correspond to various turbulent levels $\eta$. We compare the results of the semi-analytical ODE approach (thick-dashed lines) with those of numerical simulations (thin-full lines). The ballistic regimes are displayed with thin dotted lines. Asymptotic values for the two cases, ODE and simulations, are indicated by empty and filled arrows, respectively. In the MAG case only, dashed colored lines correspond to the predictions of the modified set of ODE equations (Eq.~\eqref{eq:sigx2_Dperp_tau}--\eqref{eq:Dpara_int_tau}), while the ones of the original set (Eq.~\eqref{eq:sigx2_Dperp_tau_prescription}--\eqref{eq:Dperp_int_tau_prescription}) are shown with light grey dashed lines.\label{fig:rdc_tau}}
\end{figure*}

\begin{figure*}[tbh]
\includegraphics[scale=0.48]{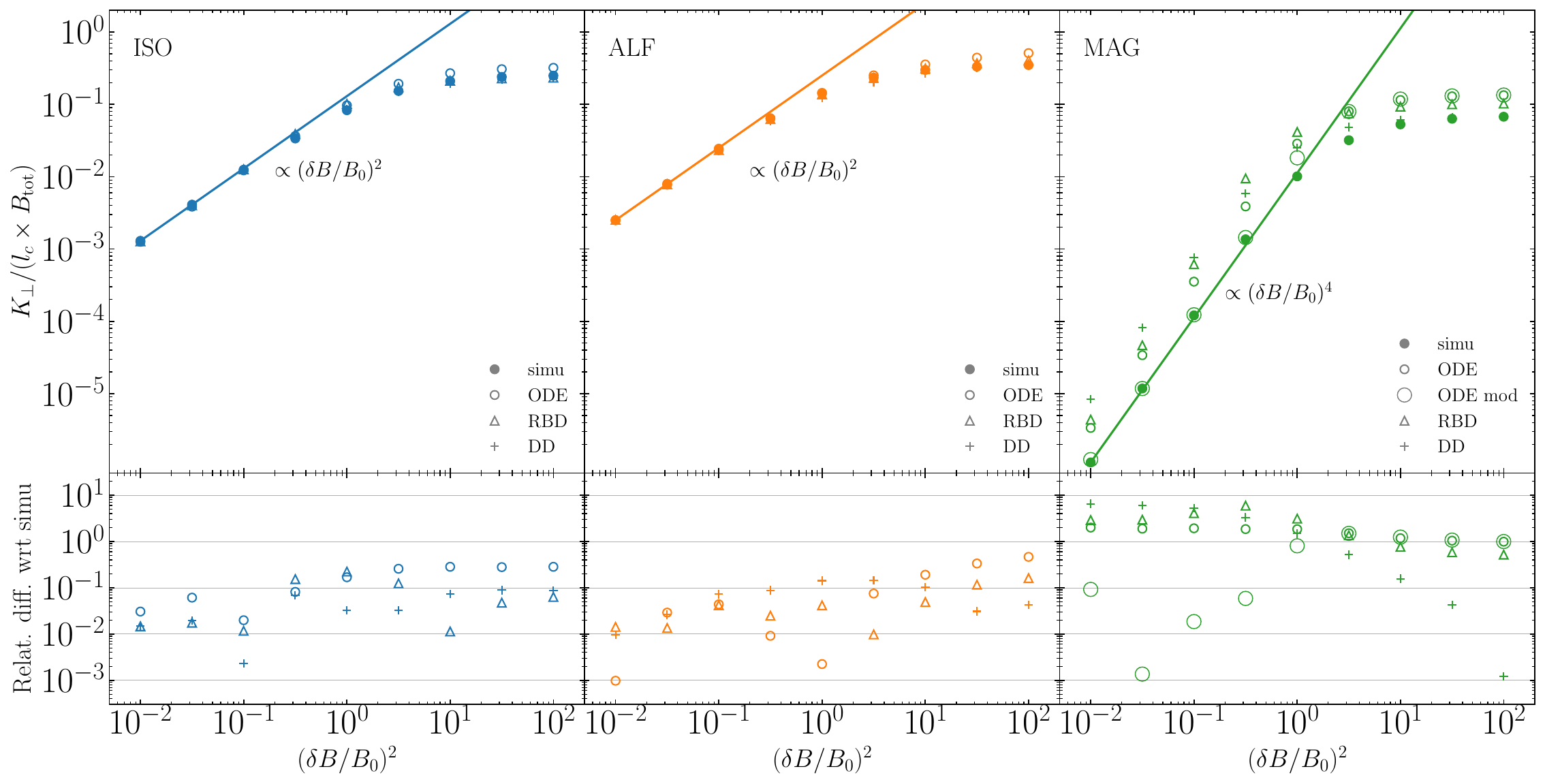}
\caption{Converged value $K_\perp$ of the perpendicular FL diffusion coefficient in the $\tau$ variable as a function of the turbulence strength. Each panel corresponds from left to right to the polarization cases: ISO, ALF and MAG. Filled circles are the values obtained from the simulations (filled arrows of Fig.~\ref{fig:rdc_tau}), empty circles denote the simulation results, empty triangles the RBD limit, and crosses the DD limit. The absolute values of the relative differences with respect to the simulation are shown with the same legend in the lower panels. In the MAG case only, large open circles correspond to the results from the modified set of ODE, Eqs.~\eqref{eq:sigx2_Dperp_tau_prescription}-\eqref{eq:Dperp_int_tau_prescription}. Solid lines show the scalings $(\delta B/B_0)^2$ and $(\delta B/B_0)^4$.} 
\label{fig:Kperp_tau_conv} 
\end{figure*}
\subsection{FLRW from simulations\label{sec:FL_sim}}
From synthetic simulations we compute the FL running diffusion coefficient across the mean field with respect to both the $s$ and the $\tau$ variable, according to the definitions in Eqs.~\eqref{eq:perp_rdc_s} and \eqref{eq:perp_rdc_tau}. The results for the $\tau$ variable are presented in Fig.~\ref{fig:rdc_tau} with a solid line whose color refers to a given turbulence level $\eta$. For completeness, we report the results for the $s$ variable in Fig.~\ref{fig:rdc_s} of App.~\ref{app:D}. Each of the three panels corresponds to one polarization case. The common trends on the evolution of the FL running diffusion coefficient can be summarized as follows:\\~\\
\textit{(i) The ballistic regime:} for small $\tau$ ($s$) values, typically $\tau\ll L_c/B_{tot}$ ($s\ll L_c$) the running diffusion coefficient increases linearly with $\tau$ ($s$). This behavior characterizes the ballistic regime, for which Eq.~\eqref{eq:perp_rdc_tau} can be approximated by:
\begin{equation}
D_\perp \approx \frac{1}{2}\left( \langle \delta B_x^2 \rangle + \langle \delta B_y^2 \rangle \right)\tau = \frac{1}{2} \left(\frac{1}{3}+\frac{2}{3}\rho\right)\; \delta B^2 \tau \;.
\label{eq:balistic_tau}
\end{equation}
The coefficient $\rho$ is defined in Eq.~\eqref{eq:def_rho} and equals $0.5$, $1$ and $0$ in the ISO, ALF and MAG cases, respectively. In the dimensionless units used in Fig.~\ref{fig:rdc_tau}, the running diffusion coefficients in the ballistic regime simply scale as $\eta$, and so saturate for small $B_0$.\\

\noindent \textit{(ii) Intermediate regime:} this regime applies at the transition from the ballistic to the diffusive regime. The general behaviors, common to the different polarization cases --- i.e the three panels of Fig.~\ref{fig:rdc_tau} --- are the following: as $\tau$ increases, the running diffusion coefficient starts softening (sub-ballistic phase), reaches a peak value at about $\tau\approx B_{\rm tot}/L_c$ ($s\approx L_c$), then decreases to a minimum (sub-diffusive phase), and exhibits an oscillatory pattern before converging to a constant value. We see that the sub-diffusive phase is steeper (i.e. $\alpha$ in Eq.~\eqref{eq:diff_subdiff} smaller) and the minimum reached lower for smaller $\eta$ values. On the contrary, for $\eta$ values close to 1 this transient fades out. The running FL diffusion coefficient corresponds to the growth rate of mean-square displacement of the ensemble of FLs. As said earlier, in the ballistic regime this growth rate is positive, and increases as $\tau$. However, for $\tau$ approaching a correlation length, FLs are bending thus the growth rate slows down, being proportional to $\tau^\alpha$, with $\alpha<1$. For larger values of $\tau$, the mean-square displacement can stall (or even decrease), causing the growth rate to drop (or even become negative). The oscillatory trend arises from the subsequent acceleration or deceleration of the transverse mean squared displacement, every correlation length (note the logarithmic scale for $\tau$ in Fig.~\ref{fig:rdc_tau}). The fact that this behavior is more pronounced for small $\eta$ or large $B_0$ values can be attributed to the relatively smaller transverse displacements in these cases. More importantly, we note drastic differences among the three polarization cases under scrutiny. Strikingly, the sub-diffusive phase is much more pronounced in the MAG case compared to the ISO and ALF cases, especially for low $\eta$ values: the relative amplitude between the peak and the converged value of the running diffusion coefficient reaches two orders of magnitude in the MAG case, while it is less than a factor 2 in the two other cases. It is difficult to explain heuristically such an order-of-magnitude difference. However this behavior can be elucidated by the correlations of the transverse component of the magnetic field discussed in the previous section: in the MAG case, the FLs moving along $B_0$ see transverse (almost) uncorrelated perturbations which tend to bring FLs back to their initial positions $(x_0,y_0)$. This is illustrated by the typical patterns of the transverse magnetic field which tend to trap field lines in the MAG case and not in the ISO and ALF cases (see Appendix~\ref{app:B}). 

In the next section we show that solving the nonlinear field line equations under Corrsin's approximation gives a good description of this transient, in all the polarization cases.  
\\
\noindent \textit{(iii) The diffusive regime:} After the sub-diffusive transient, FL running diffusion coefficients are reaching a plateau characteristic of the diffusive regime. 
We denote these converged coefficients as $K_\perp$ (see Eq.~\eqref{eq:limits}), explicitly marking their values in Fig.~\ref{fig:rdc_tau} with filled arrows and reporting them with full dots as a function of $\delta B/B_0$ in Fig.~\ref{fig:Kperp_tau_conv}. The general trend in the three polarization cases is the same: $K_\perp$ values increases as a power-law of the fluctuations' amplitude $\delta B/B_0$ for small perturbation amplitudes and flatten out above $\delta B/B_0\approx 1$. We expect such a transition since in the limit $B_0 \ll \delta B$, $B_0$ becomes irrelevant in the FLs motion, and diffusion reaches the Bohm limit $K_\perp \propto \delta B$ \citep[see discussion in][]{2015ApJ...798...59S}. In this limit, polarization loses its physical significance, as no preferred direction exists. 

Remarkably, we notice a different scaling for the ISO and ALF cases for which $K_\perp$ scales as $(\delta B/B_0)^2$, whereas for the MAG case it scales as $(\delta B/B_0)^4$. These behaviors can be explained with the following heuristic argument. For the regime $\delta B/B_0 \ll 1$ i.e. relatively small perturbation amplitude compared to the background field, the typical scattering angle of magnetic field lines is $\theta^*\approx \delta B_\perp/B_0 \propto \delta B/B_0 \ll 1$. In this regime, the typical value of the FL transverse diffusion coefficient is given by:
\begin{equation}
K_\perp \approx \frac{(\Delta r_\perp)^2}{l_{B_\perp}(\theta_*\to 0)} \approx \theta_*^2 \;l_{B_\perp}(\theta_*\to 0) \;, 
\end{equation}
with $\Delta r_\perp$ the transverse displacement at each scattering estimated as:
\begin{equation}
\Delta r_\perp \approx \theta_* \;l_{B_\perp}(\theta_*\to 0)\;.
\end{equation}
For the ISO and ALF cases, $l_{B_\perp}(\theta_*\to 0)$ tends to a constant fraction of $l_c$, and thus $K_\perp \propto \theta_*^2 \propto (\delta B/B_0)^2$. Instead for the MAG case, $l_{B_\perp}(\theta_*\to 0)\propto \theta_*^2$ (see Eq.~\eqref{eq:scaling_lBperp}), and so $K_\perp \propto \theta_*^4 \propto (\delta B/B_0)^4$. This reproduces the scaling shown Fig.~\ref{fig:Kperp_tau_conv}.
We also note that for a given $\delta B/B_0$ we always have the hierarchy: $K_\perp^{\rm MAG}<K_\perp^{\rm ISO}<K_\perp^{\rm ALF}$. From Eq.~\eqref{eq:equip} : $\delta B_\perp^{\rm MAG}<\delta B_\perp^{\rm ISO}<\delta B_\perp^{\rm ALF}$, however we show in the App.~\ref{app:E}, that the hierarchy in the diffusion coefficient $K_\perp$ still holds when enforcing $\delta B_\perp$ to be the same in the different polarization cases. Hence, the hierarchy does not only come from the different partition of $\delta B$ in the $\perp$ and $\parallel$ directions, but is also driven by the differences in $l_{B_\perp}$ discussed in Sec.~\ref{sec:aniso_lbperp}. \\      
We refer the reader to App.~\ref{app:D} for results presented for the arc length $s$ integration instead of $\tau$. The results are qualitatively very similar and, as expected, equivalent in the small $\eta$ limit.\\

\subsection{FLRW from ODE\label{sec:FL_ode_tau}}
We also predict the perpendicular running diffusion coefficients solving numerically the nonlinear FL equations under the Corrsin's and Gaussian displacements hypothesis Eq.~\eqref{eq:sigx2_Dperp_tau}-\eqref{eq:Dpara_int_tau}. The results are shown with dashed lines in Fig.~\ref{fig:rdc_tau} for the three polarization cases and the same $\eta$ values as in the simulations. Empty arrows indicate the converged value of the diffusion coefficient for ODE calculations, to be compared with the filled arrows for the simulations. In the three polarization cases, the shape of the running diffusion coefficient is well reproduced, notably for strong background fields $B_0$: the ballistic phase, the sub-diffusive phase (that the nonlinearity of the equations succeeds in reproducing) and then the diffusive plateau. The best agreement is found for the ISO and ALF cases, where the ODE results match quite closely the simulations both for the evolution of the running diffusion coefficient and the converged values. The agreement is excellent for turbulence levels up to $\eta=0.24$ and $\eta=0.76$, respectively.
Note that, in these two cases, for larger values of $\eta$ the ODE results give discrepant results from the simulations by 40 to 60\%. Among the two hypotheses—Corrsin’s hypothesis and the Gaussian displacement hypothesis—we have verified that the latter is in good agreement with the simulation results~\citep[see also][]{Snodin_2016}. Hence, the Corrsin's independence hypothesis should be challenged, using higher order correlations that are required to account for FL trapping, characteristic of such regimes dominated by the fluctuations $\delta B$. Theoretical approaches such as percolation theory~\citep{1992RvMP...64..961I} or decorrelation path method~\citep{1998PhRvE..58.7359V} are promising refinements to go beyond the Corrsin's hypothesis.
Intriguingly, focusing on small $\eta$ values in the MAG case, the ODE predictions that perform remarkably well for the ISO and ALF cases, are barely able to reproduce the ballistic regime and the first oscillations. These predictions are shown as light grey dashed lines in the Fig.~\ref{fig:rdc_tau}. 

We note that the converged values for the diffusion coefficient is never reproduced and discrepant by a factor larger than 2 from the simulations (see the residuals Fig.~\ref{fig:Kperp_tau_conv}). We believe that this mismatch arises from a limitation of the $\tau$ formalism in the MAG case. Indeed, focusing on the regime $\eta\ll 1$, the set of Eq.~\eqref{eq:Dperp_int_s}, depending on the $z$ variable, is able to reproduce remarkably well the full shape of the running diffusion coefficient in the corresponding simulations (see Fig.~\ref{fig:rdc_s} in App.~\ref{app:D}). This result is recovered when taking the limit $\sigma_z\to 0$ in Eq.~\eqref{eq:Dperp_int_tau} from which we recover the Eq.~\eqref{eq:Dperp_int_s}. Thus, we conclude that for the MAG case in the $\tau$ formalism, $\sigma_z$ seems to have a too big weight in Eq.~\eqref{eq:Dperp_int_tau}. Lacking of appropriate theory to describe FL diffusion in this case for all $\eta$, a good enough prescription is to modify Eq.~\eqref{eq:Dperp_int_tau} introducing a $\eta$ dependent term such as:
\begin{align}
{\dd \,\sigma_x^2\over \dd \tau} &= 2\, { D}_{\perp} \ , \label{eq:sigx2_Dperp_tau_prescription} \\
{\dd \,{ D}_{\perp}\over \dd \tau}   &=  \int \dd\vct{k} P_{xx}(\vct{k}){\cos(B_0 k_\parallel \tau) \exp\left( -{\sigma_x^2 k_\perp^2 \over 2}-{\eta\; \sigma_z^2 k_\parallel^2 \over 2} \right)} \ . \label{eq:Dperp_int_tau_prescription}
\end{align}
Using this prescription, the running diffusion coefficients and corresponding asymptotic values are in remarkably good agreement with the simulations, especially at low $\eta$ values. These results are shown with dashed colored lines in the MAG panel of Fig.~\ref{fig:rdc_tau}, to be compared with the dashed light grey ones. For values of $\eta$ larger than 0.5, the differences between the simulations and the ODE are similar to the ones without this prescription, the dashed light grey and colored ones overlap in Fig.~\ref{fig:rdc_tau}. The asymptotic values of these predictions are also reported on Fig.~\ref{fig:Kperp_tau_conv} and labeled \textit{ODE mod}.

\subsection{FLRW from DD and RDB \label{sec:DD_RBD}} 
As discussed in Sect.~\ref{sec:closure}, closure methods such as DD and RBD can be useful to predict the converged value of the FL diffusion coefficient. We numerically solve the system of equations \eqref{eq:Kperp_DD}-\eqref{eq:Kpar_DD} and \eqref{eq:Kperp_RBD}-\eqref{eq:Kpar_RBD} and show the results of both methods on Fig.~\ref{fig:rdc_tau}. The relative absolute differences with respect to the simulations are shown on the lower panel of the same figure. We observe that, for the ISO and ALF cases, the DD and RBD results are in good agreement with the simulations, and reproduce the shape of $K_\perp$ as a function of $(\delta B/B_0)^2$ with maximal deviations of order 10\%. Particularly, in the low–turbulence-amplitude limit, the difference reaches a few percent. In the MAG case however, even if the global trend is reproduced, the results are quite discrepant for the simulation: DD is about a factor 10 larger than the simulations at small $(\delta B/B_0)^2$, but gets closer than 10\% for values above $(\delta B/B_0)^2=10$, while the RBD results oscillate between a factor 10 and 2 above the simulation results. This mismatch is not surprising, as in the previous section we already identified discrepancies in the ODE approach—upon which the DD and RDB closures are based—when using the $\tau$ formalism.\\ 

However, the main interest of DD and RBD closures is to be able to recover analytically the scaling of $K_\perp$ in the limit of small perturbation amplitudes. In the following we show that we can recover the scaling found in the simulations, in the ISO/ALF cases but also in the MAG case using simple arguments. We will below first consider the DD closure and then the RDB one.\\

A general expression for $K_\perp$ and $K_\parallel$ in the DD model is given by Eqs. \eqref{eq:Kperp_DD}-\eqref{eq:Kpar_DD}. In the high $B_0$ limit, \citet{2016ApJS..225...20S} have shown in their section 3.3
\begin{align}
    K_\perp &= {\rho} \frac{l_c}{4} {\delta B^2 \over B_0} \;, \\ K_\parallel &= (1-\rho) \frac{l_c}{2} {\delta B^2 \over B_0} \ .
\end{align}
Note that our equations differ from theirs by a factor of two because of our definition of $l_c$. For $\rho \neq 0$ it is clear that the scaling for the dimensionless $K_\perp/(l_{\rm c} B_{\rm tot})$ diffusion coefficient is $\left(\delta B/B_0\right)^2$. In simulations, the latter is recovered in the ISO and ALF cases for which $\rho$ equals 0.5 and 1, respectively. However, for the MAG case, $\rho = 0$, a value for which the prediction of $K_\perp$ given by the above formulae fails. Nevertheless, the asymptotic scaling $\left(\delta B/B_0\right)^4$ can be derived going to second order in $K_\perp$. We first fix $\rho =0$, and get
\begin{align}
    K_\perp^{(1)} &= 0 \;, \\ K_\parallel^{(1)} &=   \frac{l_c}{2} {\delta B^2 \over B_0} \; . 
\end{align}
The index (1) indicates the first order of the calculation.
The next order in $\delta B/B_0$ is obtained by reinserting $K_\perp^{(1)}$ and $K_\parallel^{(1)}$ into Eqs. \eqref{eq:Kperp_DD} and \eqref{eq:Kpar_DD}. We find:
\begin{align}
    K_\perp^{(2)} &= \int \dd\vct{k} P_{xx}(\vct{k}) {K_\parallel^{(1)} k_\parallel^2 \over \left( K_\parallel^{(1)} k_\parallel^2  \right)^2 + B_0^2 k_\parallel^2} \\ 
    &\approx \frac{K_\parallel^{(1)}}{B_0^2} \int \dd\vct{k} P_{xx}(\vct{k})\\
    &\approx \frac{1}{6}\frac{l_c}{2} {\delta B^4 \over B_0^3}\ ,
\end{align}
where the last two lines have been obtained by retaining only the first-order terms in the limit $\delta B/B_0 \ll 1$. We thus have that the perpendicular transport is dominated by second-order effects, $K_\perp \approx K_\perp^{(2)}$, while $K_\parallel\approx K_\parallel^{(1)}$ is dominated by the first order term. We have checked that this scaling and normalization corresponds to the results found by the direct numerical integration of Eqs.~\eqref{eq:Kperp_DD}-\eqref{eq:Kpar_DD}.  In the limit of small $\eta$, the relation with the diffusion coefficient as a function of $s$ is $K_\perp = {\cal K}_\perp B_0$ so 
\begin{equation}
    {\cal K}_\perp \simeq \frac{l_c}{12}  {\delta B^4 \over B_0^4} \ .\label{eq:DD_as}
\end{equation}
This result also agrees with the scaling found in the simulations with respect to $s$, as shown in App.~\ref{app:D}.

Likewise, in the RDB closure, we can recover the MAG scaling in the limit $\delta B/B_0\ll 1$, but with a different normalization. Starting from Eqs.~\eqref{eq:Kperp_RBD}, and following section 3.5 of \citet{2016ApJS..225...20S}, by setting $\rho=0$, we can write:  
\begin{align}
    K_\perp&= \frac{\pi}{2}\int \dd k \int_{-1}^1 \dd x \,\frac{\sqrt{2}\pi^{3/2} k P(k)x^2}{B_0\sqrt{\nu'(x)}}g_{\mu,\sigma}(\chi_2)
    \label{eq:RBD_scaling_int}
\end{align}
where $g_{\mu,\sigma}$ is a Gaussian function of average $\mu=0$ and variance $\sigma=\delta B/B_0$, and $\chi_2= x/ \sqrt{2 \nu'(x)}$ with $\nu'(x)= (1+ 3 x^2)/12$. For $\rho\neq 0$, the RBD result in the limit $\sigma=\delta B/B_0\ll 1$ was obtained approximating the narrow Gaussian $g_{\mu,\sigma}(\chi_2)$ with a Dirac distribution $\delta(\chi_2)=\sqrt{2 \nu'} \delta(x)$. If we do so in the MAG case, i.e., for $\rho=0$, we see that we get $K_\perp=0$, as a first-order approximation (i.e., when $g(\chi_2) \approx \delta(\chi_2)$). However, when $\sigma\to 0$, we can perform the moment asymptotic expansion of $g_{\mu,\sigma}$ at the next order~\citep{estrada1993asymptotic}, this gives:
\begin{equation}
    g_{\mu,\sigma}(X)\approx \delta(X-\mu)+\frac{\sigma^2}{2}\delta''(X-\mu) + \dots\,
\end{equation}
with again $\mu=0$ and $\delta''$ the second derivative of the Dirac distribution. After some calculations (see App.~\ref{app:F} for more details), we recover the scaling found numerically:
\begin{equation}
    K_\perp \simeq \frac{l_c}{24} {\delta B^4 \over B_0^3}\ .\label{eq:RBD_as}
\end{equation}
We have checked that this normalization matches our numerical results for RBD in the limit $\delta B/B_0\ll 1$, and that it stands a factor 2 below the DD limit, Eq.~\eqref{eq:DD_as}.

\begin{figure*}[tbh]
\centering
\includegraphics[scale=0.45]{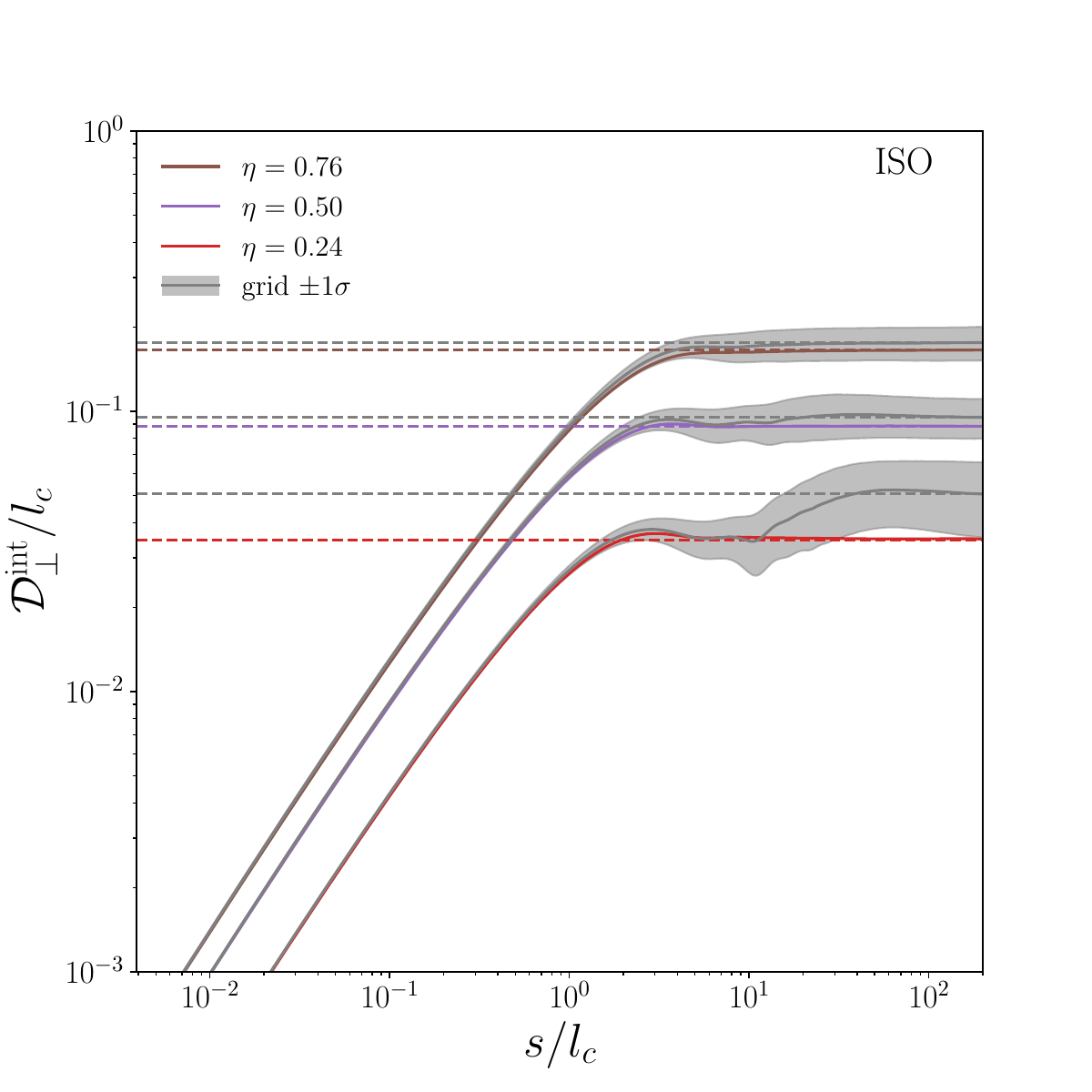}
\includegraphics[scale=0.45]{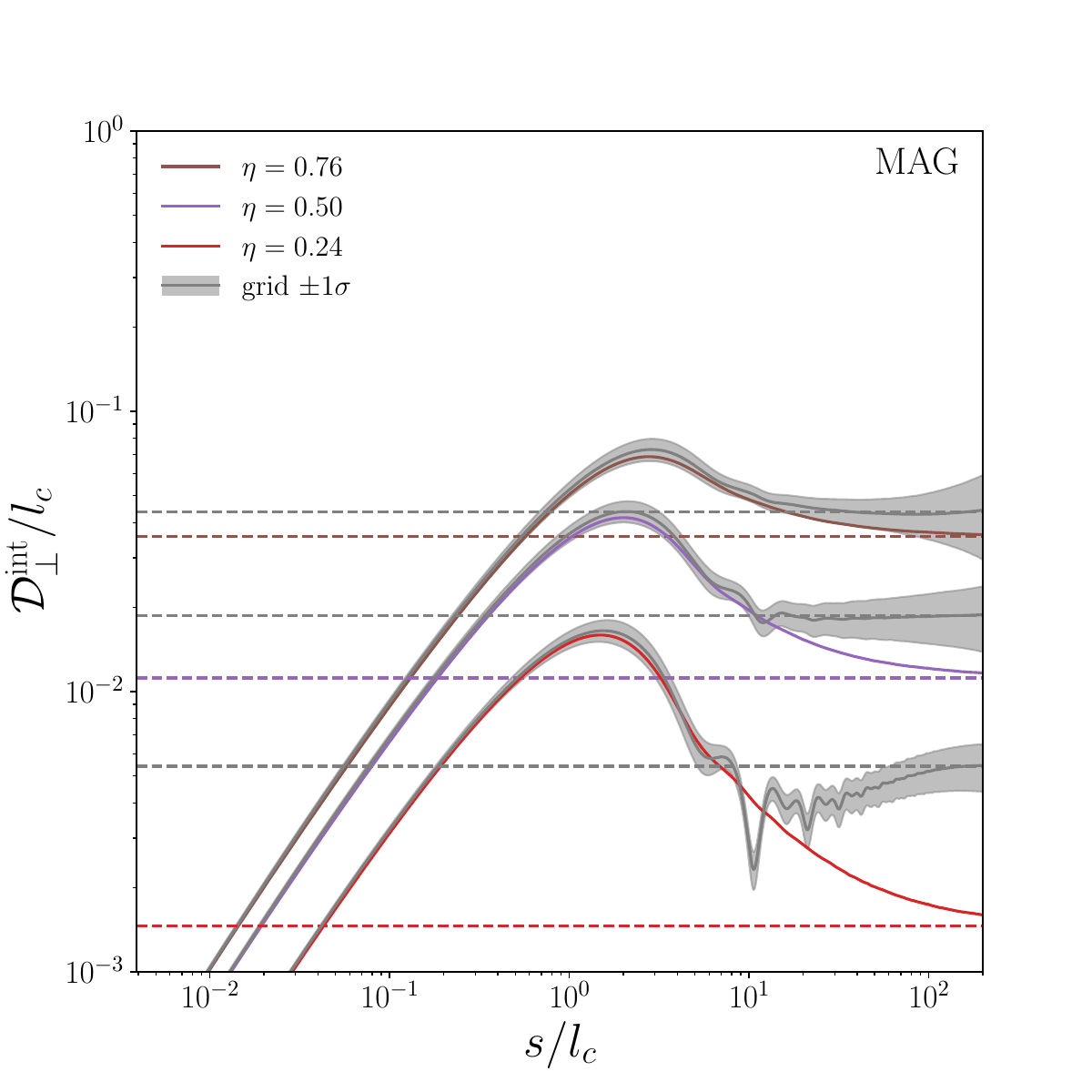}
\caption{Running FL diffusion coefficients computed using Eq.~\eqref{eq:def_int_s} in synthetic turbulence simulation using both the harmonic method (colored lines) and the grid method with $L_{\rm inj}=L_{\rm box}/2$ (gray lines). The shaded region around the gray lines shows the standard deviation estimated from ten different realizations. The asymptotic values are shown with horizontal lines. The left panel corresponds to the ISO polarization case, the right panel shows the MAG case. The discrepancy between the grid and harmonic methods, particularly evident in the MAG case and in general for small $\eta$ values, highlights the bias that can arise in computing FL diffusion when the correlation is too close to the box size.}
\label{fig:Kperp_grid_harmo} 
\end{figure*}

\section{Discussion on FL reconstruction in DNS\label{sec:V}}
Using direct numerical simulation (DNS) of MHD turbulence, several papers have investigated FL transport, mostly focused on Richardson superdiffusion~\citep{2014ApJ...784...38L, 2022MNRAS.512.2111H}, since it could impact acceleration of CR in perpendicular shocks. They investigate FL transport using particles with very small Larmor radii \citep{2013ApJ...779..140X} or directly reconstructing FL structures \citep{2024A&A...691A.149N}. As emphasized in these paper, both the converged value and the evolution of the running FL diffusion coefficient highlight key properties of the turbulence. 
Recently, \citet{2025ApJ...994L..49K} simulated forced MHD turbulence at unprecedented resolution and computed the transverse FL diffusion coefficient (see their figure~3):
\begin{equation}
    {\cal D}_{\perp}^{int} = \frac{1}{2} \frac{\langle\Delta r_\perp^2\rangle}{s}\;, \label{eq:def_int_s}
\end{equation}
which is the integrated version of the diffusion coefficient presented in this paper Eq.~\eqref{eq:def_running_s}. Interestingly, they do find a significant subdiffusive phase in the running FL diffusion coefficient, with a ratio of the maximum to the converged value of about $4$. In the following we would like to compare this behaviour with the one found for polarized synthetic turbulence using a similar fluctuation amplitude $\delta B/B_0\approx 2$.
In this exercise we are aware of the limitation of a direct comparison between synthetic turbulence and DNS, so the results should not be over interpreted. However, from the results presented in this paper, magnetosonic-type fluctuations seem to play a key role in shaping FL subdiffusion, and this raises the question of the relative importance of these fluctuations in DNS of compressible MHD turbulence.

In Fig.~\ref{fig:Kperp_grid_harmo} we display the running diffusion coefficients computed in synthetic turbulence for the ISO (upper panel) and MAG (lower panel) both with the harmonic method, shown with colors, and the grid method shown in gray \citep[for details about the  method, see][]{2020Ap&SS.365..135M}. The gray band shows the standard deviation computed from the ten realizations over which the average is taken. The diffusion coefficients are showns for three values of $\eta=0.76,0.50,0.24$ i.e. $\delta B/B_0\approx 1.78,1.0,0.56$, respectively. For the grid method we are using a $1024^3$ box size with injection length $L_{\rm inj}=L_{\rm box}/2$, which is a value commonly used in MHD turbulence simulations to get a reasonable inertial range. This leads to a coherence length $l_{\rm c}\approx L_{\rm box}/10$ for the Kolmogorov type turbulence we are using. For the two simulation methods we compute the transverse running diffusion coefficients using Eq.~\eqref{eq:def_int_s} (full lines), and display the converged value with a horizontal dashed line. 
Note that the running diffusion coefficients shown with colored lines are using the same simulations as in Appendix.~\ref{app:D}, but the definition used is Eq.~\eqref{eq:def_int_s} instead of Eq.~\eqref{eq:def_running_s}. The former tends to smooth the fluctuations and so converges more slowly than the definition from Eq.~\eqref{eq:def_running_s} \citep[for more details, see also appendix A.3 of][]{2025ApJ...992...10K}. Hence, the subdiffusive phase tends to be smoothed out and is no longer visible in the ISO case with definition Eq.~\eqref{eq:def_int_s}. In the MAG case the ratio between the maximal value and the asymptotic value of the running diffusion coefficients goes from $\approx2$ to $\approx3$ in both methods, but is systematically larger in the harmonic method. Interestingly, these values are of the same order than the factor $\approx4$ found in \citet{2025ApJ...994L..49K}. Note also that there is a systematic bias in the (running and) asymptotic values between the grid method and the harmonic method, the latter always giving lower values than the first. This bias comes from the fact that FLs do not decorrelate enough using a periodic box with $l_{\rm c}\approx L_{\rm inj}/5=L_{\rm box}/10$, while there is no such a limit in the harmonic method. Indeed, we checked that using smaller $l_{\rm c}/L_{\rm box}$, we recover the diffusion coefficient of the harmonic method. Furthermore, we note that this bias increases drastically for smaller turbulence levels (i.e. smaller $\eta$), for which the mismatch in the asymptotic value of the running diffusion coefficient can be as large as $\approx4$ for $\eta=0.24$. In this case the running diffusion coefficient seems to converge at $s/l_{\rm c}\approx 100$. However, for larger values of $s/l_{\rm c}$ (not shown in the figure) it does not. Hence, bearing in mind the limitations of a direct comparison between synthetic and MHD turbulence, we caution against a precise interpretation of reconstructions of FL diffusion coefficients in MHD boxes with not enough scale separation between $l_{\rm c}$ and $L_{\rm box}$.

\section{Conclusion\label{sec:VI}}
In this work we investigate within isotropic synthetic turbulence realizations the transport of magnetic FLs. We specifically consider cases of wave polarization consistent with the ones characterizing the main MHD modes, namely the Alfvén and magnetosonic waves. Even if synthetic turbulence can not catch all the richness of MHD turbulence which involves non-linear scale coupling and formation of intermittent structures \citep{2023PhPl...30d0502V}, our setup keeps an important property of MHD modes, and in that sense it goes beyond standard models \citep{2009ASSL..362.....S}. Hence, this work follows the efforts to go beyond simple synthetic turbulent fields \citep{2014ApJ...796...97S,  2016PhRvE..94e3109M,2016MNRAS.459.3395P,2022PhRvE.106b5307D,2024ApJS..273...11M,2024EL....14643001L,2025arXiv250518155L}, which are quite promising for isolating key properties of magnetic fluctuations and for aiding the interpretation of specific regimes of MHD turbulence revealed by direct numerical simulations. We have hereby tested three different types of synthetic turbulence models: the unpolarised isotropic variance case (ISO), the Alfvén-like polarization case (ALF) and the magnetosonic-like polarization case (MAG). The ISO case serves as a comparative framework in regards to previous studies \citep[e.g.,][and references therein]{2016ApJS..225...20S}. To that end, our results concerning field-line transport are explicitly expressed in terms of both the FL arc-length $s$ (App.~\ref{app:D}) and a rescaled version of it, $\tau$ (see Eq.~\eqref{eqn:def_tau} and Sect.~\ref{sec:IV}). \\

Our main findings are:
\begin{enumerate}
    \item The running perpendicular coefficients  show a ballistic regime $D_{\perp} \propto \tau$ (or $\propto s$) in all cases, followed by an intermediate sub-diffusive regime above $\tau \sim l_{\rm c}/B_0$ (or $s \sim l_{\rm c}$). Remarkably, the sub-diffusion is more marked in the MAG case at low $\eta$. This specific behavior can be qualitatively explained by a lesser transverse FL displacement in the MAG case induced by smaller correlations of  the transverse perturbations $\delta B_\perp$ along the parallel direction. Interestingly, simulations of forced MHD turbulence at unprecedented resolution~\citep{2025ApJ...994L..49K} also finds a significant subdiffusive phase in the transverse FL diffusion coefficient.
    \item The asymptotic diffusion regime shows remarkable difference among the three cases. In the low $\eta$ limit, the ISO and ALF cases follow quasi-linear theory, i.e. $K_\perp/(l_{\rm c} B_0) \propto {\cal K_\perp}/l_c \propto (\delta B/B_0)^2$, while the MAG case shows a different scaling, namely $K_\perp/(l_{\rm c} B_0) \propto {\cal K_\perp}/l_c \propto (\delta B/B_0)^4$. This can be well explained by the anisotropic correlations of the transverse perturbations $\delta B_\perp$. Additionally, we can also analytically recover these scalings by using the DD or RBD closures and performing a perturbative development in $\eta$ as induced by the effect of the diffusion parallel to the background magnetic.
    \item In the limit of low $\eta$, the ODE approach reproduces the simulation well in all cases, except for the MAG case, where an offset in normalization by a factor $\sim 2$ is found in the $\tau$ parametrization but not for the $s$ parameterization. We propose a prescription to correct this flaw in the theory.
    \item In the limit of large $\eta$, the ODE results do not match precisely the simulations, a trend already present in \citet{2016ApJS..225...20S}. One way to treat this effect could be to adapt the decorellation path method of \citet{1998PhRvE..58.7359V} to our cases, or to apply percolation theory~\citep{1992RvMP...64..961I,2014ApJ...791...51D}.  
    \item Overall, asymptotic diffusion coefficients obtained using specific closures methods (DD or RDB) provide a good match with simulations of ISO and ALF cases, but always display a significant offset in normalization in the MAG case.
    
\end{enumerate}

Let us recall that these results were obtained for isotropic turbulence with a simple Kolmogorov scaling. Natural extensions of this work include studying the influence of the power spectrum (its shape and slope) and exploring anisotropic turbulence \citep[see e.g.][]{2007PhPl...14k2901S}. In Paper II, where we investigate particle transport in such polarized synthetic turbulence, we show that an accurate description of the running diffusion coefficients of field lines—rather than only their asymptotic values—is essential for modeling perpendicular particle transport.

\nocite{*}

\begin{acknowledgements}
Y.G. warmly thanks Pierre Aubert for his help in implementing and optimizing the codes on GPUs, and Mathéo Gaillard for initiating work on this topic during his first year master’s project. This work has been done thanks to the facilities offered by the Univ. Savoie Mont Blanc (USMB)- CNRS/IN2P3 MUST computing center, and support from the Labex Enigmass R\&D Booster program and USMB grant NoBaRaCo. AM thanks the support from the INTERCOS IN2P3 master-project.
S.S.C. acknowledges support from the ANR grant ``MiCRO'' with reference number ANR-23-CE31-0016. 
\end{acknowledgements}

\bibliographystyle{aa}
\bibliography{biblio_MFL.bib}

\appendix
\nolinenumbers

\section{polarization tensors}\label{app:A}
The fluctuations' polarization tensor can be obtained by adopting the same method that is used in test particle simulations~\citep{2013PhPl...20b2302T}. 
First, define a unit vector $\uvct{y}'$ perpendicular to both $\vct{k} = k \uvct{k}$ and $\vct{B}_0=B_0 \uvct{z}$, $\uvct{y}' \equiv (\uvct{k} \times \uvct{z})/|\uvct{k} \times \uvct{z}|$,
where $\uvct{k}$ and $\uvct{z}$ are unit vectors in the direction of the wave vector and the background field $\vct{B}_0$. 

A general polarization vector $\vct{\xi}$ can now be obtained by rotating $\uvct{y}'$ around $\uvct{k}$ by an angle $\tilde{\alpha}$. 
If we denote the matrix for this rotation by $R$, then
\begin{align}
\vct{\xi} = R \uvct{y}' &= \uvct{k} (\uvct{k} \cdot \uvct{y}') + \cos \alpha (\uvct{k} \times \uvct{y}') \times \uvct{k} + \sin \alpha (\uvct{k} \times \uvct{y}')\nonumber\\ 
& = \frac{1}{|\uvct{k} \times \uvct{z}|} \left( \cos \tilde{\alpha}\, (\uvct{k} \times \uvct{z}) - \sin \tilde{\alpha} \big[\uvct{z} - \uvct{k} (\uvct{k} \cdot \uvct{z}) \big] \right) \, , 
\end{align}
or in component notation, $\xi_i = R_{ij} y_j'$, where $R_{ij} = \delta_{ij} \cos \tilde{\alpha} + \hat{k}_i \hat{k}_j (1- \cos \tilde{\alpha}) - \sin \tilde{\alpha}\, \varepsilon_{ijk} \hat{k}_k$,
and
\begin{equation}
\xi_i = \frac{1}{|\uvct{k} \times \uvct{z}|} \left( \cos \tilde{\alpha} \, \varepsilon_{ijk} \hat{k}_j \hat{z}_k - \sin \tilde{\alpha} \big[\hat{z}_i - \hat{k}_i (\uvct{k} \cdot \uvct{z}) \big] \right) \, . 
\end{equation}

The general polarization tensor then reads, 
\begin{align}
\langle \xi_i \xi_l \rangle = \frac{1}{|\uvct{k} \times \uvct{z}|^2} & \left\langle 
\left( \cos \tilde{\alpha} \, \varepsilon_{ijk} \hat{k}_j \hat{z}_k - \sin \tilde{\alpha} (\hat{z}_i - \hat{k}_i (\uvct{k} \cdot \uvct{z}) ) \right)\right.\nonumber \\
& \,\,\left.\left( \cos \tilde{\alpha} \, \varepsilon_{lmn} \hat{k}_m \hat{z}_n - \sin \tilde{\alpha} (\hat{z}_l - \hat{k}_l (\uvct{k} \cdot \uvct{z}) ) \right) \right\rangle . 
\end{align}
To match the notation of the standard polarization decomposition presented Eq.~\eqref{eq:polarization_vector}, we introduce $\alpha=\tilde{\alpha}-\pi/2$, and rewrite the polarization tensor:

\begin{align}
\langle \xi_i \xi_l \rangle = \frac{1}{|\uvct{k} \times \uvct{z}|^2} & \left\langle 
\left( \sin \alpha \, \varepsilon_{ijk} \hat{k}_j \hat{z}_k + \cos \alpha (\hat{z}_i - \hat{k}_i (\uvct{k} \cdot \uvct{z}) ) \right)\right. \\
& \,\, \left.\left( \sin \alpha \, \varepsilon_{lmn} \hat{k}_m \hat{z}_n + \cos \alpha (\hat{z}_l - \hat{k}_l (\uvct{k} \cdot \uvct{z}) ) \right) \right\rangle . 
\end{align}

\subsection*{Isotropic turbulence}

For isotropic turbulence, this needs to be averaged over $\alpha$, 
\begin{align}
\langle \xi_i \xi_l \rangle = \frac{1}{2} \frac{1}{|\uvct{k} \times \uvct{z}|^2} 
& \left\{ \varepsilon_{ijk}\, \hat{k}_j \hat{z}_k\, \varepsilon_{lmn}\, \hat{k}_m \hat{z}_n\right. \nonumber \\
& \quad+\, \left. \Big(\hat{z}_i - \hat{k}_i (\uvct{k} \cdot \uvct{z}) \Big) \Big(\hat{z}_l - \hat{k}_l (\uvct{k} \cdot \uvct{z}) \Big) \right\}\;. 
\end{align}
The first term can be expanded to $\varepsilon_{ijk} \hat{k}_j \hat{z}_k \varepsilon_{lmn} \hat{k}_m \hat{z}_n = \delta_{il} + \hat{k}_i (\uvct{k} \cdot \uvct{z}) \hat{z}_l + \hat{z}_i \hat{k}_l (\uvct{k} \cdot \uvct{z}) - \hat{k}_i \hat{k}_l - \delta_{il} (\uvct{k} \cdot \uvct{z})^2 - \hat{z}_i \hat{z}_l$,
resulting in 
\begin{align}
\langle \xi_i \xi_l \rangle &= \frac{1}{2} \frac{1}{|\uvct{k} \times \uvct{z}|^2} \left( \delta_{il} - \hat{k}_i \hat{k}_l - \delta_{il} (\uvct{k} \cdot \uvct{z})^2 + \hat{k}_i \hat{k}_l (\uvct{k} \cdot \uvct{z})^2 \right) \\
&= \frac{1}{2} \left( \delta_{il} - \hat{k}_i \hat{k}_l \right) \, ,
\end{align}
that is the usual polarization tensor for isotropic turbulence~\citep{1982tht..book.....B}. 
In spherical coordinates, where 
\begin{equation}
\uvct{k} = \left( \begin{matrix} \sqrt{1 - \chi^2} \cos \varphi \\ \sqrt{1 - \chi^2} \sin \varphi \\ \chi \end{matrix} \right) \, ,
\end{equation}
with $\chi \equiv \cos \theta$, this rewrites as
\begin{equation}
\langle \xi_i \xi_l \rangle = {{\frac{1}{2}}} 
\left( \begin{matrix}
	1 - {\cal S}\cos^2\varphi			& -{\cal S}\cos \varphi\sin\varphi		& - \chi\sqrt{{\cal S}}\cos \varphi \\
	-{\cal S}\cos\varphi\sin\varphi		& 1 -{\cal S}\sin^2\varphi			& - \chi \sqrt{{\cal S}}\sin\varphi \\
	- \chi\sqrt{{\cal S}}\cos\varphi 		& -\chi\sqrt{{\cal S}}\sin\varphi 		& {\cal S}
\end{matrix} \right)\,, \label{app:tensor_iso}
\end{equation}
where we introduced ${\cal S}=1-\chi^2$ for short.

\subsection*{Alfvènic polarized turbulence}
The polarization vector $\vct{\xi}$ for Alfvènic waves is perpendicular to both, $\uvct{k}$ and $\uvct{z}$, that is $\tilde{\alpha} = 0\; [\pi]$ or $\alpha=\pi/2 \; [\pi]$, or $\vct{\xi} \equiv \uvct{y}' = (\uvct{k} \times \uvct{z})/|\uvct{k} \times \uvct{z}|$.
Then, the polarization tensor thus reduces to
\begin{align}
\langle \xi_i \xi_l \rangle & =  \frac{1}{|\uvct{k} \times \uvct{z}|^2}
\left( \varepsilon_{ijk}\, \hat{k}_j \hat{z}_k\, \varepsilon_{lmn}\, \hat{k}_m \hat{z}_n\,  \right)\nonumber \\
& = \frac{1}{|\uvct{k} \times \uvct{z}|^2} 
\left( \delta_{il}(1 - \hat{k}_z^2) + \hat{k}_z(\hat{k}_i \hat{z}_l + \hat{z}_i \hat{k}_l) - \hat{k}_i \hat{k}_l - \hat{z}_i \hat{z}_l \right)\,,
\end{align}
which in spherical coordinates rewrites as
\begin{equation}\label{Eq:TMS}
\langle \xi_i \xi_l \rangle = 
\left( \begin{matrix}
	 1 - \cos^2 \varphi				&  - \cos \varphi \sin \varphi		& 0 \\
	 - \cos \varphi \sin \varphi		&  1- \sin^2 \varphi				& 0 \\
	0 	& 0 		& 0
\end{matrix} \right) \,. 
\end{equation}
For the case of axi-symmetric turbulence, $\langle \xi_x \xi_x \rangle = \langle \xi_y \xi_y \rangle = 1/2$.

\subsection*{Magnetosonic polarized turbulence}

The polarization vector $\vct{\xi}$ for magnetosonic waves is perpendicular to both, $\uvct{k}$ and $\uvct{y}'$, that is $\tilde{\alpha} = \pi / 2$, equivalently $\alpha = 0\; [\pi]$ or $\vct{\xi} \equiv \uvct{k} \times \uvct{y}' = [\uvct{k} \times (\uvct{k} \times \uvct{z})]/|\uvct{k} \times \uvct{z}| = [\uvct{k} (\uvct{k} \cdot \uvct{z}) - \uvct{z}]/|\uvct{k} \times \uvct{z}|$.
The polarization tensor is then  
\begin{equation*}
\langle \xi_i \xi_l \rangle = \frac{1}{|\uvct{k} \times \uvct{z}|^2} \left( \hat{k}_i \hat{k}_l (\uvct{k} \cdot \uvct{z})^2 - (\hat{k}_i \hat{z}_l + \hat{z}_i \hat{k}_l) (\uvct{k} \cdot \uvct{z}) + \hat{z}_i \hat{z}_l \right) \, . 
\end{equation*}
In spherical coordinates this rewrites as
\begin{equation}\label{Eq:TMS}
\langle \xi_i \xi_l \rangle = 
\left( \begin{matrix}
	 \chi^2\cos^2\varphi				&  \chi^2\cos \varphi\sin\varphi		& -\chi\sqrt{{\cal S}}\cos\varphi \\
	 \chi^2\cos\varphi\sin\varphi		&  \chi^2\sin^2\varphi				& -\chi\sqrt{{\cal S}}\sin\varphi \\
	-\chi\sqrt{{\cal S}}\cos\varphi 	& -\chi\sqrt{{\cal S}}\sin\varphi 		& {\cal S}
\end{matrix} \right)\,. 
\end{equation}
And for axi-symmetric turbulence, $\langle \xi_x \xi_x \rangle = \langle \xi_y \xi_y \rangle = \chi^2 / 2$. 
\section{Visualisation of FL and transverse fluctuations in polarised turbulence \label{app:B} }

\begin{figure*}[tbh]
\centering
\includegraphics[scale=0.54]{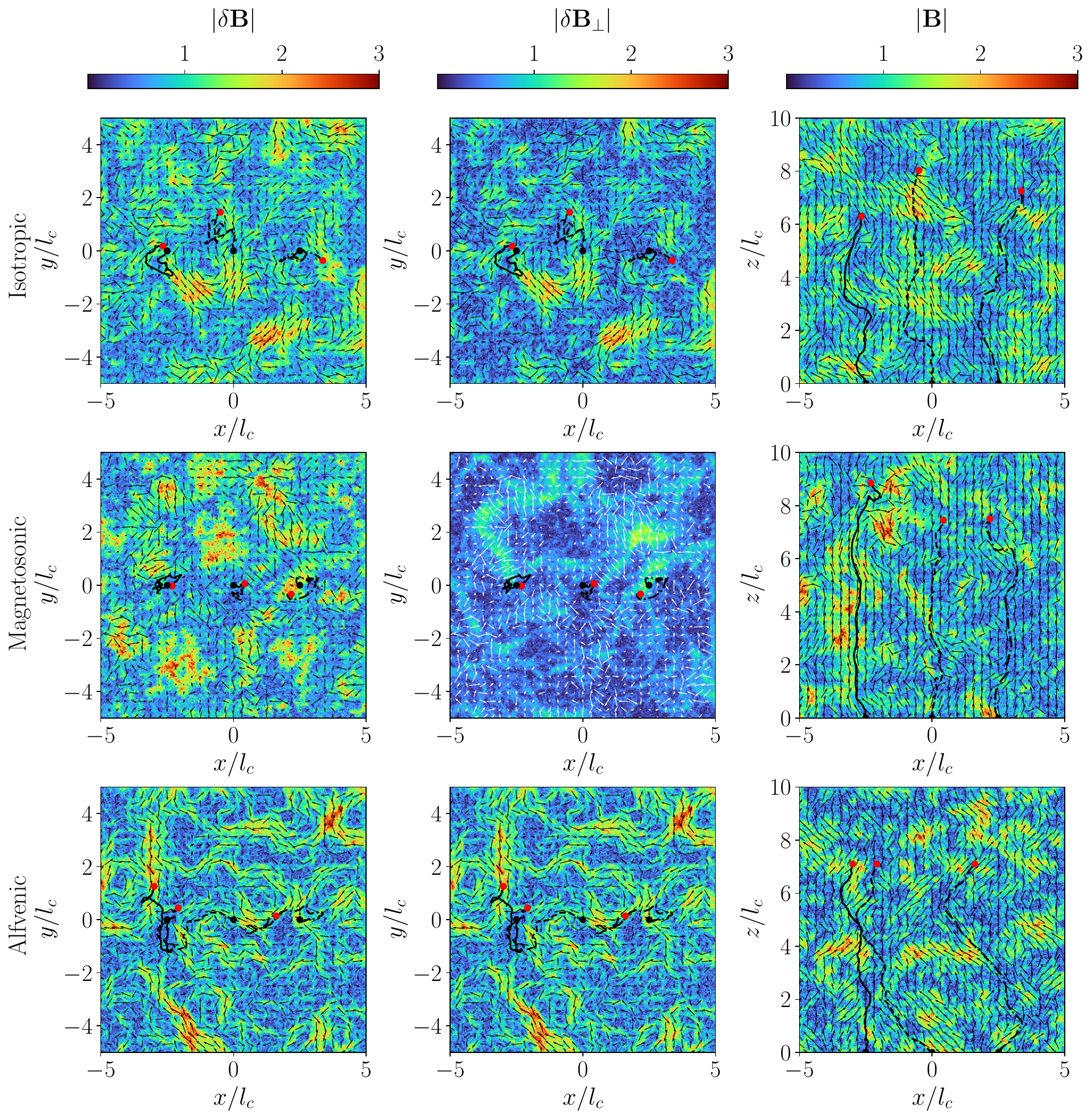}
\caption{2D maps of ${\bf B}$, $\boldsymbol{\delta}{\bf B}$ and  $\boldsymbol{\delta}{\bf B_{\perp}}$ magnitudes in harmonic simulation for plane $(x, y, z=0)$ in first and second column panels and plane $(x, y=0, z)$ for third column. From first to third row, panels show isotropic, magnetosonic, and Alfvénic polarization, respectively, with the same turbulence level $\eta=0.5$. Arrows show orientation of the local magnetic field. Three example field lines are projected onto the 2D maps. Large black dots indicate initial coordinates, large red dots indicate final coordinates of integration.}
\label{fig:2Dmap} 
\end{figure*}
In this Appendix we highlight the structures that emerge in the B-field from the different polarization configurations. In Fig.~\ref{fig:2Dmap} we show 2D color maps of the magnetic field in our synthetic turbulence simulations ($\eta=0.5$), where each of the three rows corresponds to a different turbulent field characterized by a specific polarization (ISO, ALF or MAG). The color indicate the magnitude of the magnetic field: either the full norm, $||\vct{B}||=||B_0\vct{z}+\delta\vct{B}||$, or the transverse norm $||\delta \vct{B}_\perp||=(\delta B_x^2 + \delta B_y^2)^{1/2}$. The first two columns display the transverse plane ($x$, $y$) at $z=0$, and the last column shows the vertical plane ($x$, $z$) at $y=0$. Arrows are superposed to represent the transverse magnetic field vector ($\delta\vct{B}_\perp$) in the two first planes and the total magnetic field ($\vct{B}=\vct{B}_0+\delta\vct{B}$) in the last plane. Three field lines (full, dashed, dash-dot black lines) have been reconstructed and projected onto the different sections (the black dots represent the starting points for the FL reconstruction, while the red dots represent the end points reached after integrating up to $s= 10\, l_c$). The FLs are the same for each polarization (or row), only the viewpoint changes.\\
The goal of this plot is to give some more qualitative support to the drastically different behaviours of FLs in the three polarization configurations. 
Firstly, by comparing the magnitude maps in the first two rows we can appreciate the fact that, for instance, Alfvènic turbulence has no parallel component whereas in the magnetosonic case, the parallel perturbation is dominant. We can also notice the different properties of the patterns that naturally emerge within each polarization: regions of larger fluctuations amplitude form more filamentary ``structures'' that are aligned with the local direction of the fluctuating magnetic-field $\delta\vct{B}_\perp$ for Alfv\'enic polarization, while for magnetosonic polarization these large-amplitude ``structures'' are more clumpy and their pattern is mostly perpendicular to the direction of the magnetic field fluctuations (and especially with respect to the perpendicular component $\delta\vct{B}_\perp$ --- see middle panel); for the isotropic polarization it looks like a mixture of both. We can clearly see the different behaviors of field lines according to the polarization, which supports Fig.~\ref{fig:field_line_wandering_3D} and the different displayed transverse diffusion coefficients. Indeed, for a given $\eta$, Alfv\'enic turbulence has larger $\delta B_\perp$ which are correlated over larger distances in the vertical direction (larger $l_{B_\perp} (\theta^*\to 0)$). Hence, field lines excursions are larger in the transverse plane, and consequently $D_{\perp}$ value is higher. Finally, we can see that field lines generally tend to follow the displayed field orientation as well as following the different structures, especially in the Aflvénic case. However, there are some regions where the reconstructed FLs do not agree with the local field orientation in that plane or do not follow the shape of local magnetic structures. This is simply due to a projection effect, i.e., to the fact that off-plane magnetic perturbations cannot be displayed onto the plane over which FLs have been projected.

\section{Statistics of curvatures}\label{app:C}
\begin{figure*}[tbh]
\includegraphics[scale=0.5]{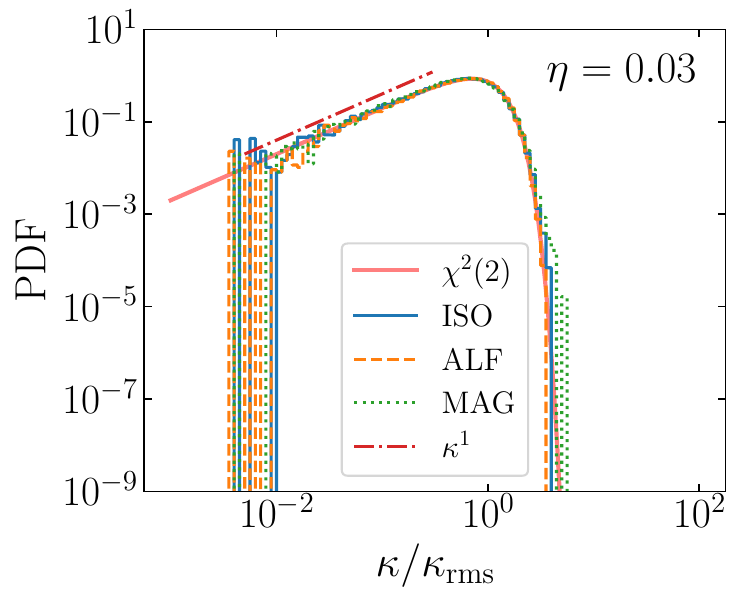}
\includegraphics[scale=0.5]{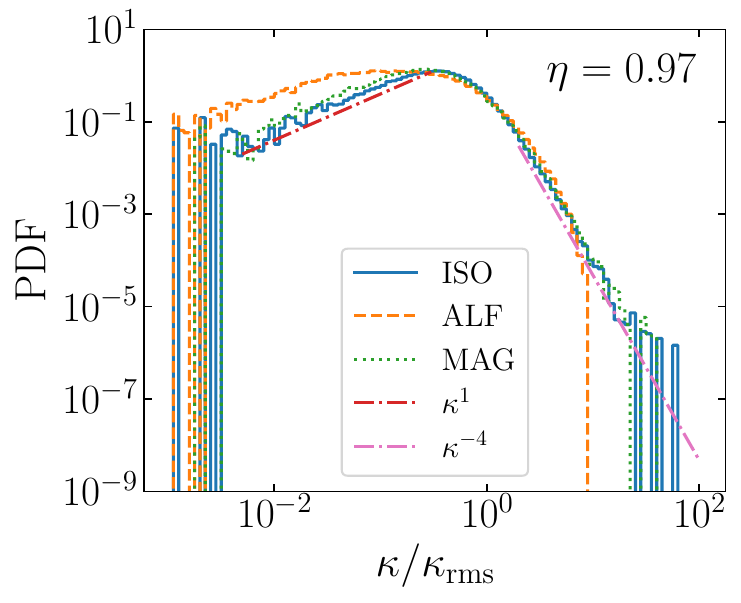}
\includegraphics[scale=0.5]{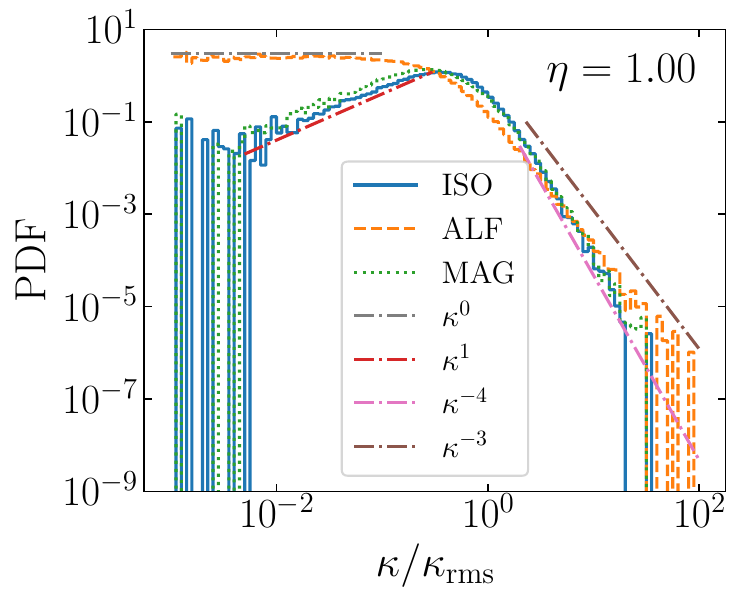}
\caption{Distribution of the curvature for the three polarization configurations used in the paper: isotropic (ISO), Alfvènic (ALF) and magnetosonic (MAG). From left to right, with increasing turbulence level.} 
\label{fig:bending}
\end{figure*}

In this Appendix we compare the distribution of magnetic-field line curvatures obtained from synthetic isotropic turbulence simulations with different polarization. The curvature of a magnetic FL is defined as 
\begin{equation}
    \kappa=\frac{|\vct{b}\times(\vct{b}\cdot\nabla \vct{B})|}{B}=|\vct{b}\cdot\nabla \vct{b}|\;. \label{eq:defcurv}
\end{equation}
We compute the normalized histograms of $\kappa$, normalized by $\kappa_{rms}\equiv\sqrt{\langle \kappa^2 \rangle}$, from the simulations of the three polarization configurations (ISO, ALF, MAG). They are shown in Fig.~\ref{fig:bending}, for three turbulence levels. 

We can see that at low turbulence level (left panel), the PDF of the curvature for the three polarization cases are very similar. They show a tail at low-$\kappa$ values proportional to $\kappa$, while they fall exponentially at high-$\kappa$ values. We can explain this behaviour approximating $\kappa$ for small-amplitude fluctuations (i.e. small $\eta$ values), which at first order in the perturbation writes  
\begin{equation}
\kappa\approx\frac{1}{B_0}|\partial_z \delta \vct{B}_\perp | \;.\label{eq:curvature}
\end{equation}
In the axisymmetric turbulence considered in this paper, the gradients $\partial_z {\delta B_x}$ and $\partial_z {\delta B_y}$ can be considered independent and are similarly distributed, following Gaussian distributions with zero mean and same variance. Thus, the random variable $2\kappa^2/\kappa_{\rm rms}^2$ follows a $\chi^2(n=2)$, which after changing variable, gives the PDF:
\begin{equation}
p(\hat{\kappa}\equiv\kappa/\kappa_{\rm rms})=2\hat{\kappa}\exp(-\hat{\kappa}^2).
\end{equation}
This PDF, shown with a red line in Fig.~\ref{fig:bending}, is reproducing quite well the three histograms. Note that $\kappa_{\rm rms}$ diverges for infinitely large dynamical range ($\kappa_{\rm rms} \propto k_{max}^{2/3}$). However for finite $k_{max}$, it is finite but different in the three cases. In the weak turbulence limit we can compute their values analytically and find the following ratios: $\kappa_{\rm rms}^{\rm ALV}=(5/4)\,\kappa_{\rm rms}^{\rm ISO}$ and $\kappa_{\rm rms}^{\rm MAG}=(3/4)\,\kappa_{\rm rms}^{\rm ISO}$. This highlights the fact that, on average, Alfv\'enic fluctuations are the ones that produce more consistent bending of magnetic-FLs in the small-amplitude limit, due to their specific polarization.\\

The middle and right panels of Fig.~\ref{fig:bending} show a case at high turbulence level $\eta=0.97$ and an extreme case where $\eta=1$. In these cases, the curvature PDF of the ALF case differs from the ISO and MAG cases which are similar. For high-curvature values, the PDF falls with a power-law tail $\hat{\kappa}^{-4}$ in the ISO and MAG cases, whereas it still falls exponentially (for $\eta=0.97$) or as $\hat{\kappa}^{-3}$ (for $\eta=1$) in the ALF case. Furthermore, in the ALF case only, the low-curvature tail starts deviating from $\propto\kappa$ scaling at $\eta=0.97$ and completely flattens for $\eta=1$.

\begin{figure}[tbh]
\centering
\includegraphics[scale=0.65]{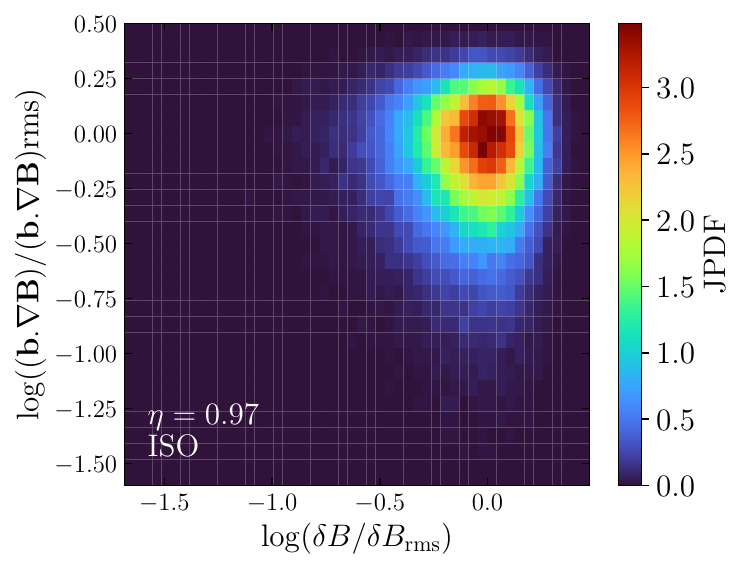}
\caption{Joint PDF from synthetic turbulence showing the correlation between the numerator of $\kappa = |{\bf b}\times({\bf b}.\nabla){\bf B})|/B$ and $\delta B$.} 
\label{fig:correlations} 
\end{figure}

In the $\delta B \gg B_0$ regime, neither of the numerator $X\equiv|\vct{b}\times(\vct{b}\cdot\nabla \vct{B})|$ nor the denominator $Y^{-1}\equiv B$ of $\kappa=X\,Y$ (see Eq.~(\eqref{eq:defcurv})) can be simplified as previously. However, $X$ and $Y$ can be considered as independent variables, since physically $X$ is set by the smallest scales of the turbulence (or high-$k$ values) with the strongest gradients and $Y^{-1}$ is fixed by the largest scales (or low-$k$ values) with largest amplitudes. In synthetic turbulence these scales are uncorrelated. We illustrate this property by plotting the joint PDF in Fig.~\ref{fig:correlations}. 

Let's consider first the $Y$ distribution. In the case $\eta=1$, $B_0$ vanishes, thus $Y^{-2}= \delta B_x^2 +\delta B_y^2 +\delta B_z^2 $. In the ISO case, $\delta B_i$ follows independent Gaussian distribution of same variance, which is not the case for MAG in which $\langle\delta B_x^2\rangle = \langle\delta B_y^2\rangle = \frac{1}{4}\langle\delta B_z^2\rangle$ (see Eq.~\eqref{eq:equip}). In practice, for these two cases the PDF of $Y^{-2}$ is well approximated by a distribution proportional to $\chi^2(n=3)$ (Note that in the MAG case, it should technically follow a generalised $\chi^2$ distribution which happens to be well approximated by a $\chi^2(n=3)$ up to some numerical factor). After changing variable, in the ISO and MAG cases, the PDF of Y is such that:
\begin{equation}
    p_Y(y) \propto y^{-4}\exp\left(-\frac{1}{2y^2}\right)\;,
\end{equation}
which for large values of Y is decreasing as a powerlaw $y^{-4}$, and has an exponential cut off toward small values. In the ALF case, magnetic fluctuations on the $z$-axis are vanishing, and $Y^{-2}$ follows a $\chi^2$ distribution with only two degrees of freedom. In that case: 
\begin{equation}
    p_Y(y) \propto y^{-3}\exp\left(-\frac{1}{2y^2}\right)\;.
\end{equation}

Let's now consider the distribution of $X$, which can be written as $B |(\vct{b}\cdot\nabla)\vct{b}| =|(\vct{b}\cdot\nabla)\vct{B} - ((\vct{b}\cdot\nabla)\vct{B}\cdot\vct{b})\vct{b}|$.
This expression shows that the numerator can be expressed by the vector $(\vct{b}\cdot\nabla)\vct{B}$ to which we remove the component parallel to $\vct{b}$. It means that, in a rotated coordinate system where $\vct{b}$ is one of the basis vectors, this vector has only two independent components. We have checked that these two components are normally distributed, hence, for the same reasons as the $Y$ variable, the distribution of $X^2$ is proportional to a $\chi^2(n=2)$ distribution in the ISO and MAG cases, while for the ALF case it is proportional to a $\chi^2(n=1)$ distribution. After changing variable we get the distribution of $X$ in the ISO and MAG cases:
\begin{align}
         p_X(x)\propto x\exp\left(-\frac{x^2}{2}\right)\;,
\end{align}
and in the ALF case:
\begin{align}
    p_X(x)\propto \exp\left(-\frac{x^2}{2}\right)\;.
\end{align}

Using the fact that $X$ and $Y$ are independent random variables, the probability distribution of $\kappa$ can be written as:
\begin{align}
    p(\kappa)\propto \int dx\;\int dy\; p_X(x) p_Y(y) \delta(xy-\kappa)  \;.
\end{align}
Using this expression, it can easily be shown that the behaviour of $p(\kappa)$ at low $\kappa$ values is driven by the heavier tail of $p_X(x)$. Similarly, at high-$\kappa$ values, the scaling of $p(\kappa)$ is set by the power-law tail of $p_Y(y)$. Hence, for the ISO and MAG cases we recover the tails $\propto \kappa$ and $\propto \kappa^{-4}$ at low and high $\kappa$ values, respectively. We thus recover the tail $\propto \kappa^{-4}$ characteristic of a Gaussian random-phase field \citep[see e.g.][]{Schaefer01012012,2024EL....14643001L,2025arXiv250518155L}. In the ALF case at $\eta=1$ we recover the tails $\propto \kappa^0$ and $\propto \kappa^{-3}$ for low and high-$\kappa$ values, respectively. The distribution of $\kappa$ in the ALF case at $\eta=0.97$ exhibits a behaviour that is a mixture between the $\eta=0.03$ and $\eta=1$ distributions.
The high curvature tail is thus different from the one found in 3D MHD, for which based on numerical simulations, the curvature PDF was observed to decrease as $\kappa^{-2.5}$ \citep{2019PhPl...26g2306Y}. Interestingly, the low-curvature tail can be similar and increases as $\kappa$. The explanation of these slopes goes within the same lines as the ones provided for synthetic turbulence above, except that the curvature can now be expressed as $\kappa=f_n/\delta B^2$, with the magnitude of the tension force normal to the field lines $f_n$, with two independent degrees of freedom. It is argued that low-$\kappa$ values are associated with small force magnitude, hence $f_n^2$ follows a typical $\chi^2(n=2)$ distribution which leads to the PDF of $\kappa\propto f_n$ having low-$\kappa$ tail proportional to $\kappa$. Instead, the high-$\kappa$ tail would be associated with low $\delta B^2$ values, following a $\chi^2(n=3)$, leading to the PDF of $\kappa\propto\delta B^{-2}$ having a high-$\kappa$ tail proportional to $\kappa^{-2.5}$. Remarkably, these asymptotic behaviours have been observed in the solar wind turbulence \citep{2020ApJ...893L..25B}. Attempts to recover such a distribution in synthetic turbulence are presently pursued \citep{2024EL....14643001L,2025arXiv250518155L}.

\begin{figure*}[tbh]
\centering
\includegraphics[scale=0.4]{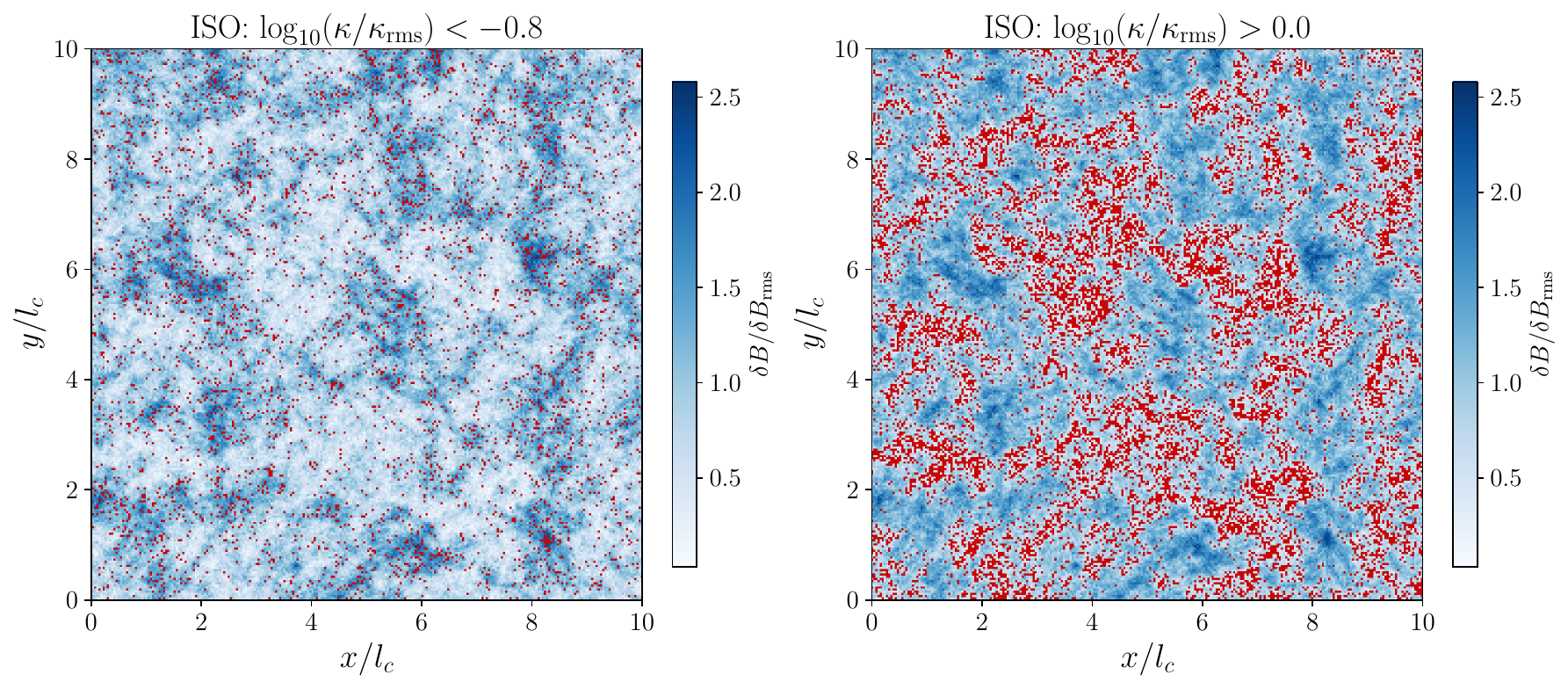}
\includegraphics[scale=0.4]{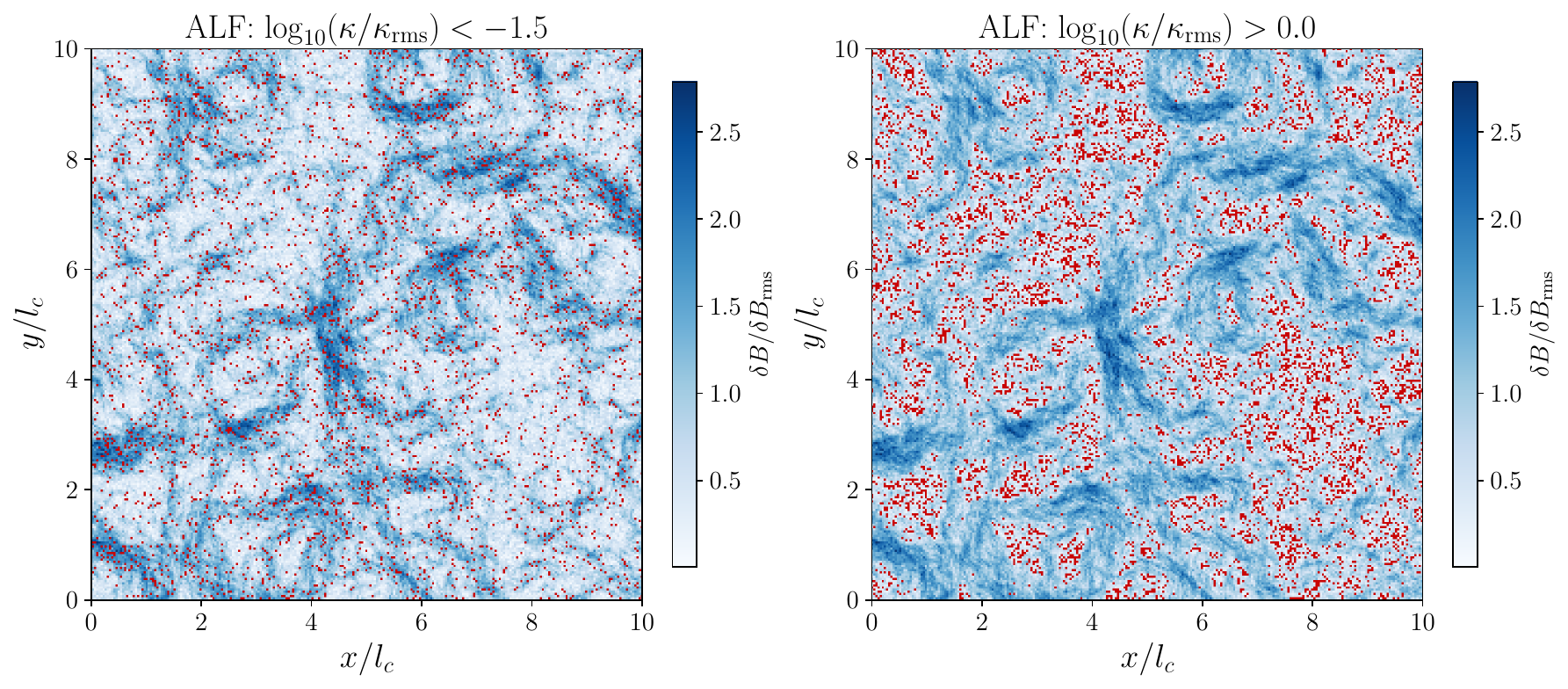}
\includegraphics[scale=0.4]{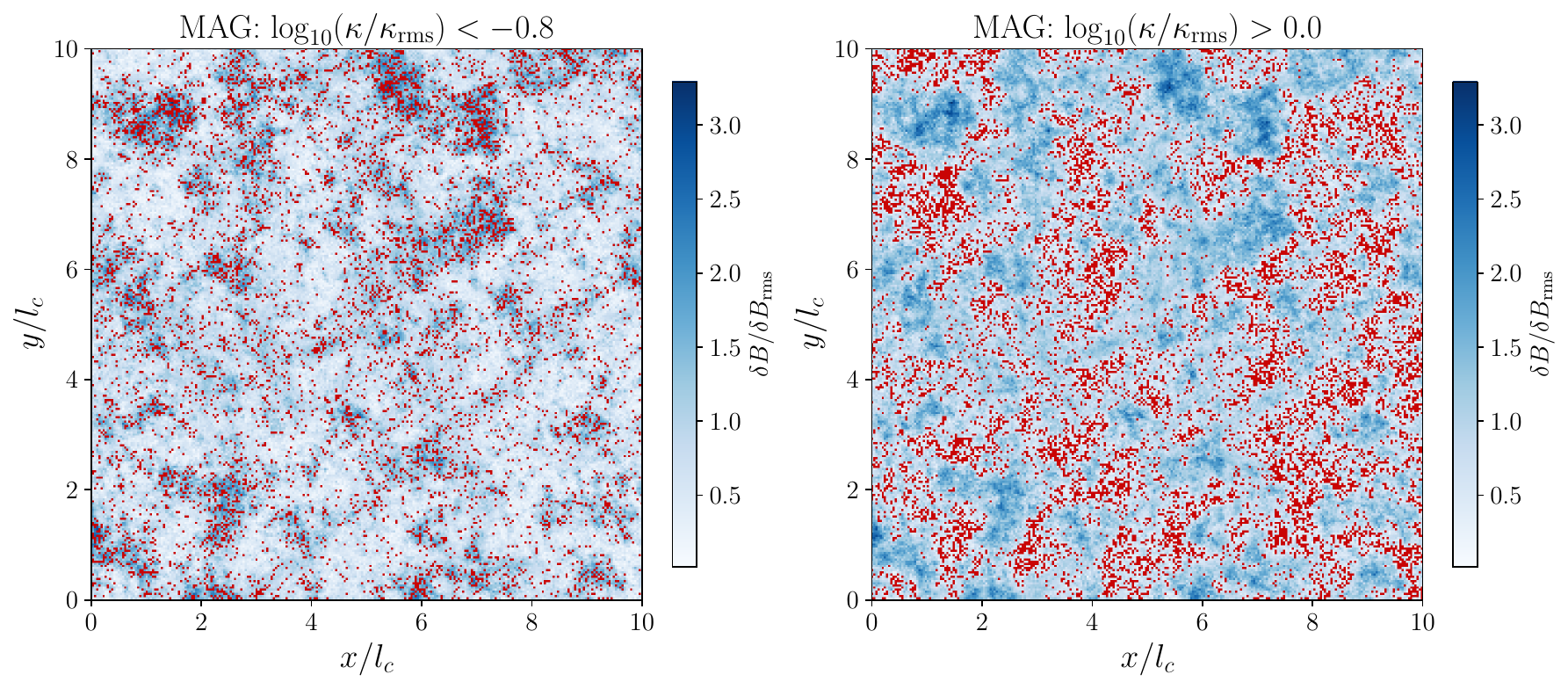}
\caption{Color map of magnetic amplitude in a square region of side $20l_c$, in the three polarization configurations: ISO (top), ALF (middle), MAG (bottom). For the three cases the turbulence level is $\eta=1$, i.e., $\vct{B}_0=0$. Regions with low (left column) and high (right column) curvature are superimposed with red color. The threshold values have been chosen to display a similar density of points. High curvature values populate regions of low magnetic field. However, low curvature values spread quite homogeneously. }
\label{fig:contour_map} 
\end{figure*}

Although they do not exactly share  the same PDF for $\kappa$, an interesting common feature of the synthetic and MHD turbulence is the spatial correlation of high curvature values with low magnetic field regions. This is a direct result of the above discussion and is illustrated in Fig.~\ref{fig:contour_map}, to be compared with Fig.~3 of ref.~\citet{2019PhPl...26g2306Y}. High curvature regions are correlated with large scales structures (right panel) because they are enhanced in regions with small magnetic amplitudes, driven by large scales perturbations. On the other hand, low curvature regions are mostly uncorrelated with these structures (left panel) because they depend on the distribution of the gradient of the magnetic field, i.e. small scales perturbations.

\section{Diffusion in $s$ vs $\tau$}\label{app:D}
This appendix gathers the results of FL diffusion studied with respect to the arc-length $s$, complementary to Sect.~\ref{sec:IV} which present the analysis performed with respect to the $\tau$ parameter. Note that, in the limit of large guide field, i.e. $B_0\to \infty$ (or, equivalently, $\eta\ll 1$), the simulation results for the two parameterizations are equivalent since one is just a constant rescaling of the other: $s\approx B_0 \tau$. Thus, the conclusions for $s$ parameterization in that limit remain the same, so we do not repeat them here. However, we highlight the differences in the simulations between the approaches based on $s$ and $\tau$ formalism when $\eta$ is close to 1. Furthermore, we show that, in the $s$ variable, the semi-analytical results based on the ODE approach and associated closures give a much better agreement with the simulations in the MAG case. This motivates the prescription proposed in Eq.~\eqref{eq:sigx2_Dperp_tau_prescription}.\\ 

\begin{figure*}[tbh]
\centering
\includegraphics[scale=0.325]{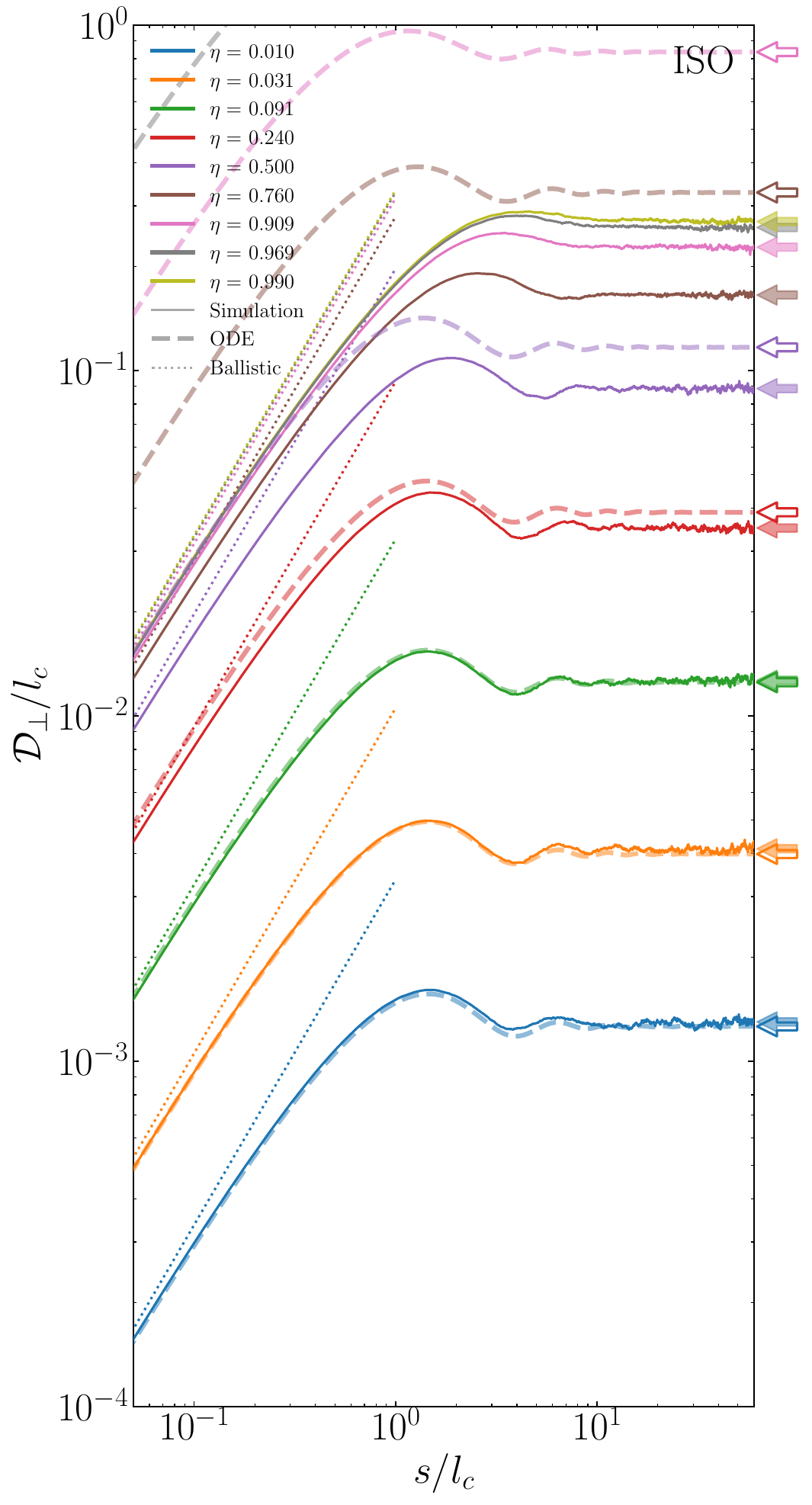}
\includegraphics[scale=0.325]{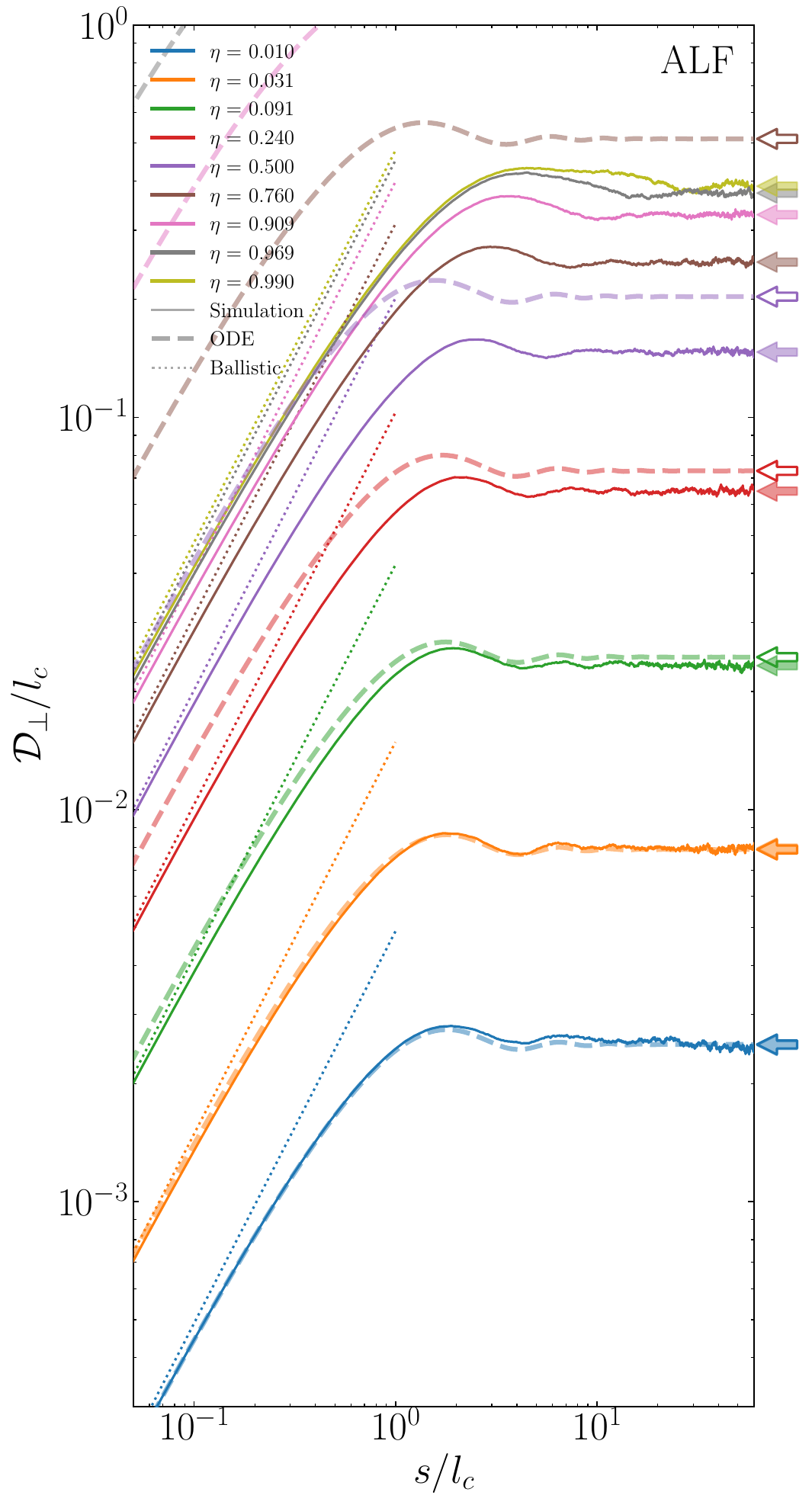}
\includegraphics[scale=0.325]{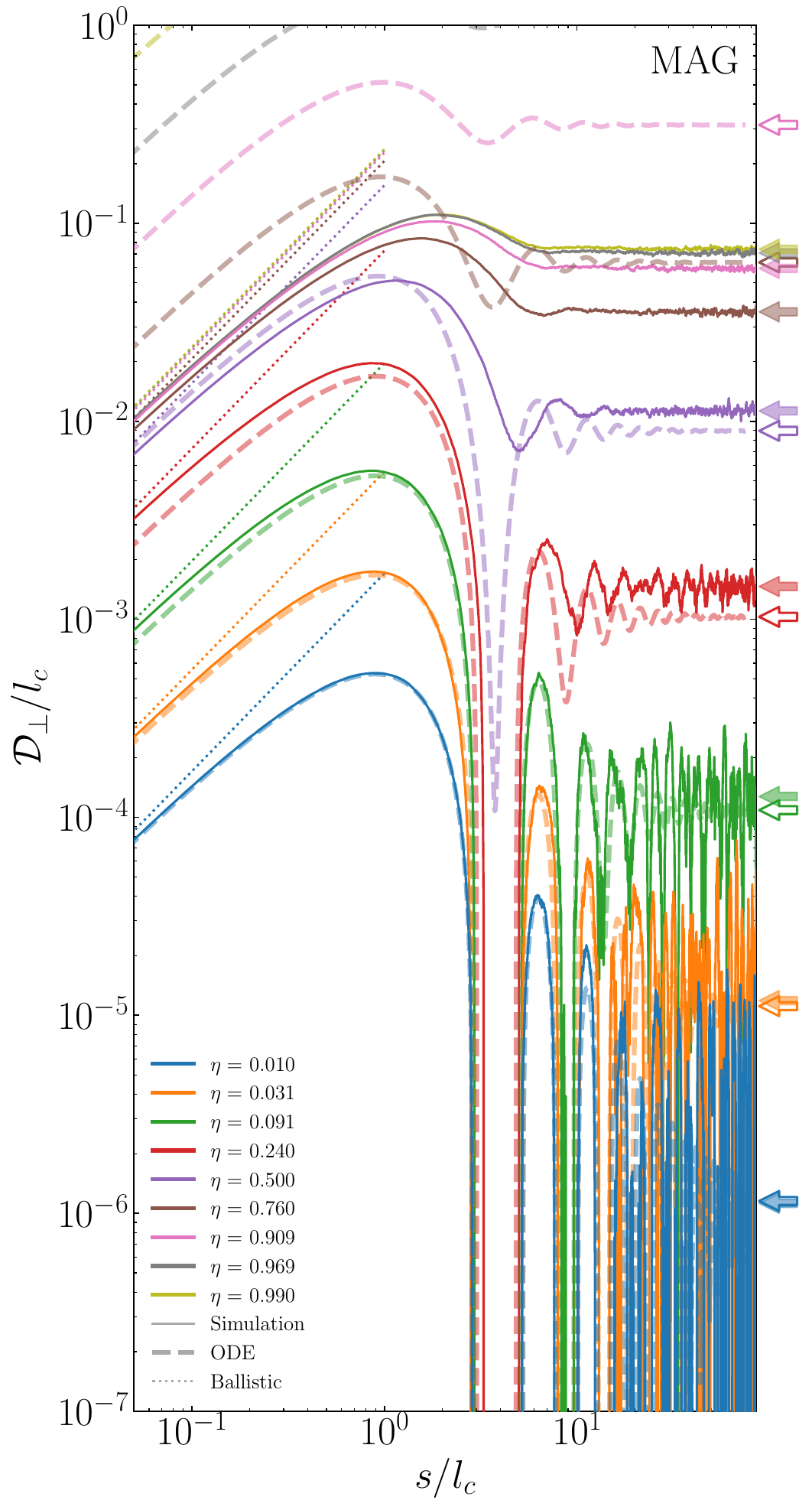}
\caption{Running FL diffusion coefficient ${\cal D}_{\text{FL}}$ as a function of $s$ for isotropic turbulence. The different colors correspond to various turbulent level $\eta$. We compare the results of the semi-analytical ODE approach (thick-dashed lines) with those of numerical simulations (thin-full lines). The ballistic regimes are displayed with thin dotted lines. Asymptotic values for the two cases, ODE and simulations, are indicated by empty and filled arrows, respectively. \label{fig:rdc_s}} 
\end{figure*}

\subsection*{FLRW from simulations}
In Fig.~\ref{fig:rdc_s} we present the results of the running FL diffusion coefficient as a function of $s$, see Eq.~\eqref{eq:perp_rdc_s}. The numerical setup is exactly the same as described in Sect.~\ref{sec:III}, only the integration of the FL is different. We recover the three regimes described in the paper:\\

\textit{(i) The ballistic regime:} for small values of $s$, typically $s\ll L_c$ the running diffusion coefficient increases linearly with $s$. This behaviour characterizes the ballistic regime, for which Eqs.~\eqref{eq:perp_rdc_s} can be approximated by :
\begin{align}
{\cal D}_\perp & \,\approx\, \frac{1}{2} \left\langle\frac{ \delta B_x^2  +  \delta B_y^2 }{B_{tot}^2} \right\rangle s\nonumber \\
& \underset{\delta B\to 0}{\approx}\, \frac{1}{2}\left( \frac{\langle \delta B_x^2 \rangle + \langle \delta B_y^2 \rangle}{B_0^2} \right) s
 \,\approx\,\frac{1}{2}\eta \;\left(\frac{1}{3}+\frac{2}{3}\rho\right) \; s\;,
\label{eq:balistic_s}
\end{align}
with $\rho$ being defined as Eq.~\eqref{eq:def_rho} and depending on the polarization configuration. The factorization proposed in the last two terms is only valid for small enough $\delta B$, in practice for $\eta<0.1$. In Fig.~\ref{fig:rdc_s} we show with dotted lines this regime, computed from the first equality of Eq.~\eqref{eq:balistic_s}, i.e. without approximation.\\

\textit{(ii) Intermediate regime:} the transition from the ballistic to the diffusive regime is quite similar to the one observed with respect to the $\tau$ variable, and so the comments written Sect.~\ref{sec:IV} also apply here. The main difference we note is the slightly more pronounced sub-diffusive phase at high $\eta$ values. This is easily visible in the MAG case.\\

\textit{(iii) The diffusive regime:}
FL running diffusion coefficients are reaching a plateau characteristic of the diffusive regime. 
We denote these converged coefficients as ${\mathcal K}_\perp$ (see Eq.~\eqref{eq:limits}), explicitly marking their values in Fig.~\ref{fig:rdc_s} with filled arrows and reporting them with full dots as a function of $\delta B/B_0$ in Fig.~\ref{fig:Kperp_s_conv}. As expected, the scaling of ${\mathcal K}_\perp$ with respect to $\delta B/B_0$ is the same as ${K}_\perp$ in the limit of small $\eta$ values. Since in this limit, $s$ is just a re-scaling of $\tau$ ($s\approx B_0\tau$), the difference of scaling between (ISO, ALF) and MAG cases can be explained by the same heuristic argument developed Sect.~\ref{sec:FL_sim}. At large $\eta$ values, ${\cal K}_\perp$ also reaches a constant value.

\subsection*{FLRW from ODE}
In the same way as in Sect.~\ref{sec:FL_ode_tau} for the $\tau$ variable, we can numerically solve the coupled ODE system (Eqs.~\eqref{eq:sigx2_Dperp_s}-\eqref{eq:Dperp_int_s}) derived Sect.~\ref{sec:qlt}, as a function of $z$, the coordinate along the $B_0$ field. The coordinate $z$ is a good approximation of $s$ for $\delta B \ll B_0$ ($\eta\ll 1$). Thus, we expect the results of the ODE system to reproduce well the simulations only in that limit. A detailed comparison between $s$ and $z$ parameterizations can be found in Sect.~3.4 in   \citet{2025ApJ...992...10K}. The solutions of the ODE system are presented with dashed colored lines in Fig.~\ref{fig:rdc_s} for the three polarization cases, ISO, ALF and MAG. As expected, we see that in all three cases, the agreement between the ODE results and the simulations is excellent at small $\eta$ values, and starts to deteriorate for $\eta > 0.1$. The position of the converged value is indicated by an open arrow on the right side of Fig.~\ref{fig:rdc_s} and by open circles in Fig.~\ref{fig:Kperp_s_conv}. In both figures, we see that the values of ${\cal K}_{\perp}$ predicted by the ODE at large $\eta$ do not plateau as in the simulations, but instead follow a shallower scaling with $\delta B / B_0$, as expected applying quasi-linear theory outside its validity regime \citep[see][]{2015ApJ...798...59S}. Note that, in the MAG case using the $s$ variable, the agreement between the ODE and the simulations at small $\eta$ values is much better than it was initially with the $\tau$ variable (dashed light grey lines in Fig.~\ref{fig:rdc_tau}). As discussed in the paper, this mismatch may arise from a limitation of the $\tau$ formalism in the MAG case and motivated us to introduce the prescription (with an additional $\eta$ term) in Eqs.~\eqref{eq:sigx2_Dperp_tau_prescription}--\eqref{eq:Dperp_int_tau_prescription}, so that at small $\eta$ values we recover the results of the ODE system given by Eqs.~\eqref{eq:sigx2_Dperp_s}--\eqref{eq:Dperp_int_s}.
 
\subsection*{DD and RBD closures}
\begin{figure*}[tbh]
\centering
\includegraphics[scale=0.42]{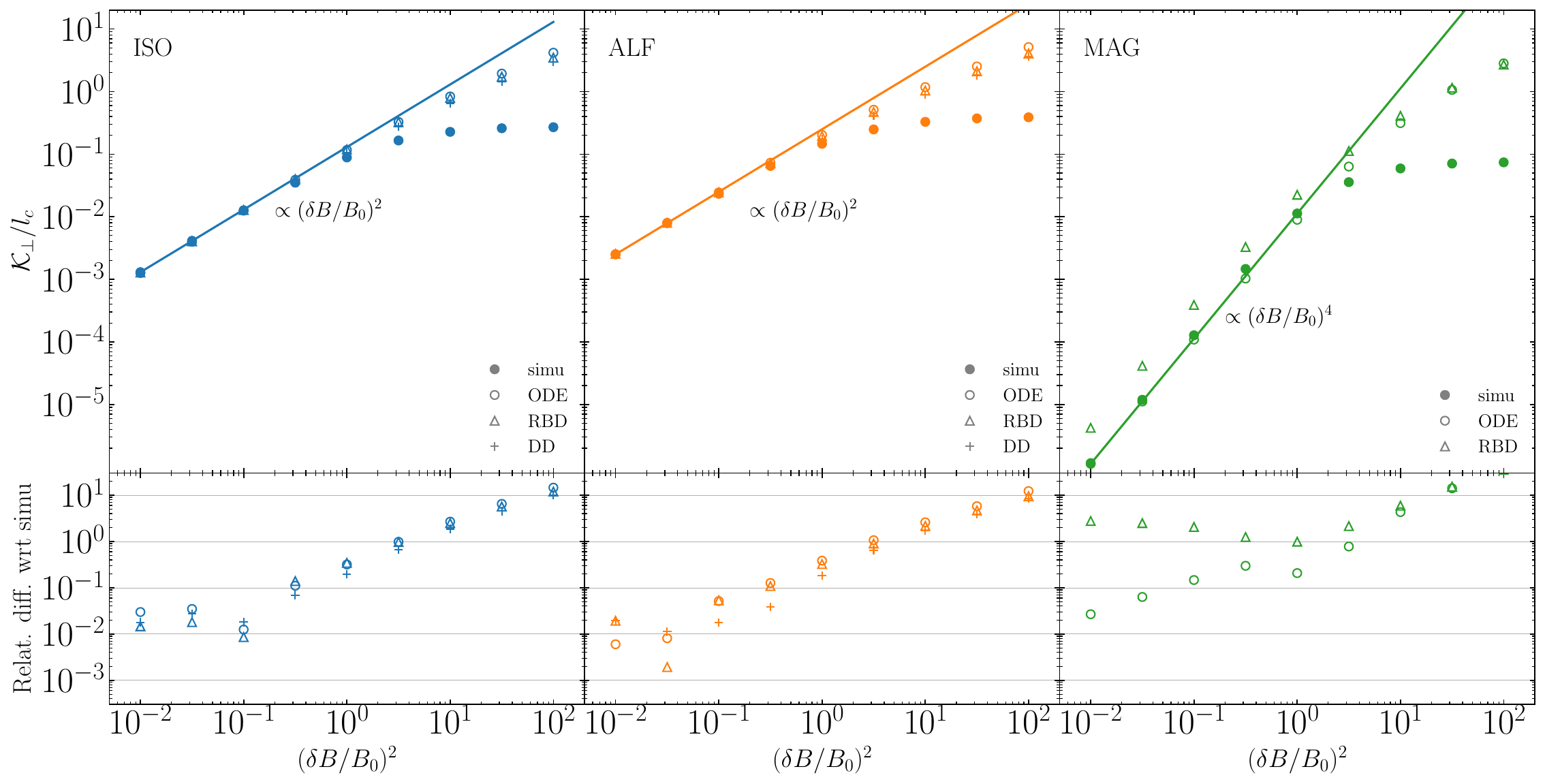}
\caption{Converged value ${\cal K}_\perp$ of the perpendicular FL diffusion coefficient in the $s$ variable as a function of the turbulence strength. Each panel corresponds from left to right to the polarization cases: ISO, ALF and MAG. Filled circles are the values obtained from the simulations (filled arrows of Fig.~\ref{fig:rdc_tau}), empty circles denote the ODE results, empty triangles the RBD limit, and crosses the DD limit. The relative differences with respect to the simulation are shown with the same legend in the lower panels. Note that the results from the DD closure are not shown for the MAG case because it was not converging numerically.} 
\label{fig:Kperp_s_conv} 
\end{figure*}

The system of Eqs.~\eqref{eq:sigx2_Dperp_s}--\eqref{eq:Dperp_int_s} can be solved introducing a closure relation. Likewise in Sect.~\ref{sec:closure}, we try the two closure relations -- the \textit{diffusive decorrelation} (DD) and the \textit{random ballistic decorrelation} (RDB) -- previously considered in the literature. The DD closure  writes $\sigma_x^2= 2{\cal K_\perp} z$ and leads to the following equation:
\begin{eqnarray} {\cal K}_{\perp} & =  \displaystyle\frac{1}{{B}_{0}^{2}} \int \dd{\vct{k}} {P}_{{xx}}({\vct{k}})\displaystyle \frac{{\cal K}_{\perp}{k}_{\perp}^{2}}{{({\cal K}_{\perp}{k}_{\perp}^{2})}^{2}+{k}_{\parallel}^{2}}\;. \label{eq:Kperp_DD_s}
\end{eqnarray}
For the RBD case, we use $\sigma_x^2= \langle \delta B_x^2 \rangle \;z^2 /B_0^2 $ which leads to 
\begin{eqnarray}
{\cal K}_{\perp}= \frac{1}{{B}_{0}^{2}} \displaystyle \frac{1}{2}\int \dd{\vct{k}}\,\sqrt{\displaystyle \frac{\pi }{{\nu }_{2}}}{P}_{{xx}}(\vct{k}){e}^{-\tfrac{{B}_{0}^{2}{k}_{\parallel}^{2}}{4{\nu }_{2}}} \ ,\label{eq:Kperp_RBD_s}
\end{eqnarray}
with ${\nu }_{2}\equiv \langle {\delta B}_{x}^{2}\rangle {k}_{\perp}^{2}$/2. We numerically solve Eqs.~\eqref{eq:Kperp_DD_s} and \eqref{eq:Kperp_RBD_s}. The results are shown in Fig.~\ref{fig:Kperp_s_conv}, to be compared with the simulations and the ODE. Note that in the MAG case, the DD results are not displayed because we encountered numerical convergence issues in the small $\eta$ regime. In the ISO and ALF cases, they show good agreement with the ODE results, but not with the simulations, as expected. For the MAG case, the RDB closure manages to recover the scaling $(\delta B/B_0)^4$ at small $\eta$ values, but not the correct normalization, which is off by a factor of approximately 3. We have checked that in the limit of large $B_0$, the results are identical to the ones given by Eq.~\eqref{eq:Kpar_DD} and \eqref{eq:Kpar_RBD}, in the three polarization cases.

\section{Impact of $\delta B_\parallel$ on the FL transverse diffusion\label{app:E}}
\begin{figure*}[tbh]
\centering
\includegraphics[scale=0.45]{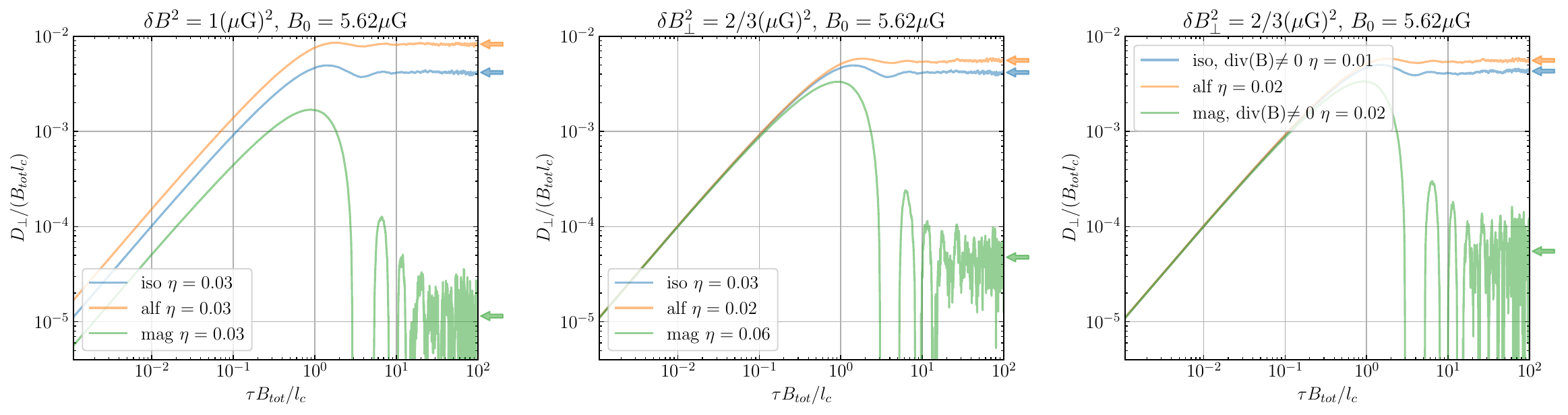}
\caption{Transverse running diffusion coefficients $D_{\perp}(\tau)$ for different polarizations from harmonic simulations at $B_0 = 10^{0.75} \approx 5.62 \rm{\mu G}$. The arrows on the right hand side of each plot point to the asymptotic value. Left panel: same $\eta = \delta B^2/(\delta B^2 + B_0^2)$ for each polarization. Middle panel: at fixed $B_0$, $\delta B$ is adjusted to reach the same $\delta B_{\perp}$ for each polarization. Right panel: same configuration as the middle panel, but we set to 0 the parallel component of the perturbed magnetic field in the magnetosonic case ($\delta B_\parallel =0$), hence ${\rm div}(B)\neq 0$.}
\label{fig:impact_Bz_Bperp_0.75} 
\end{figure*}

\begin{figure*}[tbh]
\centering
\includegraphics[scale=0.45]{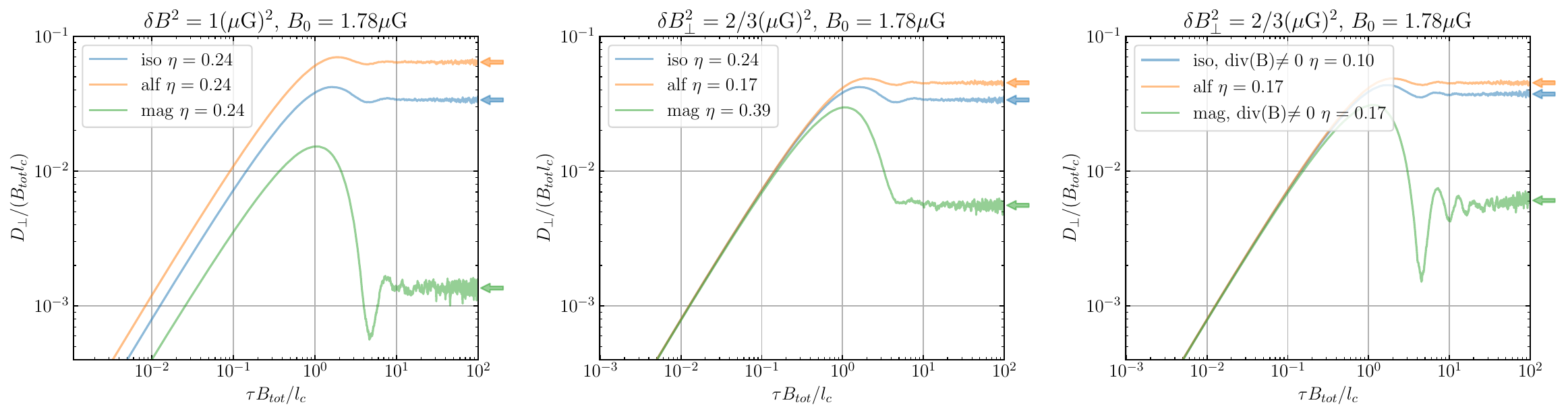}
\caption{Same legend as Fig.~\ref{fig:impact_Bz_Bperp_0.25} but for $B_0 = 10^{0.25} \approx 1.78 \,\rm{\mu G}$.}
\label{fig:impact_Bz_Bperp_0.25} 
\end{figure*}

\begin{figure*}[tbh]
\centering
\includegraphics[scale=0.45]{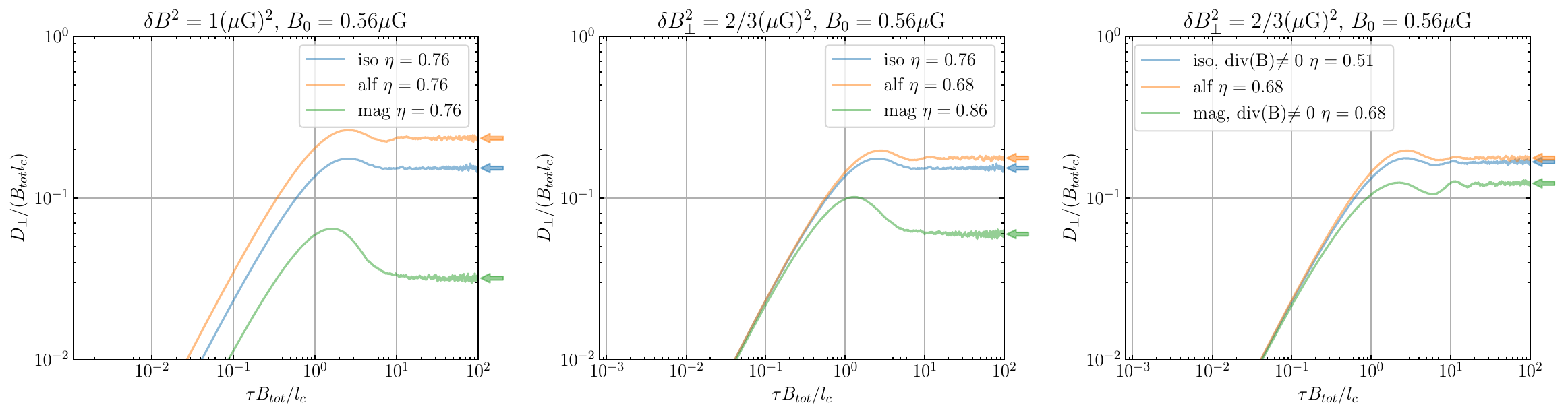}
\caption{Same legend as Fig.~\ref{fig:impact_Bz_Bperp_-0.25} but for $B_0 = 10^{-0.25} \approx 0.56\, \rm{\mu G}$.}
\label{fig:impact_Bz_Bperp_-0.25} 
\end{figure*}

In Sect.~\ref{sec:IV} we found that the transverse excursions of the FLs (and so the diffusion) is drastically reduced in turbulence with magnetosonic-like (MAG) polarization, compared to the other cases (ISO and ALF). In the paper, we ague that this difference mostly come from the anisotropy in the correlations of $\delta\vct{B}_\perp$, through $l_{B_\perp}$ defined Eq.~\eqref{eq:lperp}. In this Appendix we study the importance of another effect that contributes reducing the transverse diffusion, namely the fact that in the MAG case $\delta B_\perp^2=\frac{1}{3}\delta B^2$, while it is larger and equal to $\frac{2}{3}\delta B^2$ in ISO and $\delta B$ in ALF (see Eq.~\eqref{eq:equip}). We study the impact of this \textit{projection effect} varying the level of transverse perturbation amplitudes, and recomputing the diffusion coefficient. The results are presented Fig.~\ref{fig:impact_Bz_Bperp_0.75}, Fig.~\ref{fig:impact_Bz_Bperp_0.25} and Fig.~\ref{fig:impact_Bz_Bperp_-0.25} for different $B_0$ values, $10^{0.75},10^{0.25}$ and $10^{-0.25}\;\mu$G. In each figure, the left panels present FL diffusion coefficients for the three polarization cases at the same turbulence level $\eta$. In the middle panels, the turbulence level $\delta B$ have been adjusted to get the exact same $\delta B_{\perp}$ in each polarization. In the right panels $\delta B_{\perp}$ is the same as in the middle panels, but in the ISO and the MAG cases, we fix $\delta B_\parallel =0$ (consequently, we artificially produce a non-solenoidal field, $\boldsymbol{\nabla}\cdot{\bf B} \neq 0 $). Let us comment the evolution of the diffusion coefficients: form the left to the middle panels the turbulence level is modified to match the ISO transverse perturbations, so $\delta B_{\text{ALF}}$ is decreased and $\delta B_{\text{MAG}}$ is increased, and so $D_{\perp, ALF}$ decreases and $D_{\perp, MAG}$ increases. Here we note that, for the same $\delta B_\perp$, the running diffusion coefficients—and hence their converged values—become closer. Moreover, they converge even more closely as $B_0$ decreases, as illustrated from Fig.~\ref{fig:impact_Bz_Bperp_0.75} to Fig.~\ref{fig:impact_Bz_Bperp_-0.25}. However, there is still a significant difference between the MAG versus the ISO and ALF cases, which highlight again the importance of the anisotropic correlations of $\delta B_\perp$. Finally, to gauge the impact of $\delta B_\parallel$ on the transverse transport, we fix $\delta B_{\parallel}=0$, while keeping $\delta B_\perp$ constant in the last panels. We see that the running diffusion coefficients get closer and almost merge for the ISO and ALF cases when $B_0\lesssim \delta B$ , see Fig.~\ref{fig:impact_Bz_Bperp_-0.25}. On the other hand, when $B_0$ gets larger (with $B_0\gtrsim  \delta B$) the differences with $\delta B_{\parallel}\neq 0$ vanish (Fig.~\ref{fig:impact_Bz_Bperp_0.75}), showing as expected that $\delta B_{\parallel}$ perturbations do not impact transverse diffusion in that limit.
This observation further motivated us to introduce the prescription in Eq.~\eqref{eq:sigx2_Dperp_tau_prescription}, reducing the impact of $\delta B_{\parallel}$ in the MAG case to better match the simulation results.

\section{APPENDIX F: RBD approximation for large $B_0$ \label{app:F}}

As mentioned in the Sect.~\ref{sec:DD_RBD}, from the Gaussian distribution of mean $\mu$ and variance $\sigma$, we can perform the moment asymptotic expansion when $\sigma\to 0$, \citep{estrada1993asymptotic} and find :
\begin{equation}
    g_{\mu,\sigma}(X)=\delta(X-\mu)+\frac{\sigma^2}{2}\delta''(X-\mu)+o(\sigma^3)\,
\end{equation}
Keeping only the two first terms of this expansion, and given that $\mu=0$, we get:  
\begin{equation}
    g_{\mu,\sigma}(\chi_2(x))\approx\delta(\chi_2(x))+\frac{\sigma^2}{2}\delta''(\chi_2(x))\,,
\end{equation}
with $\chi_2(x) = \sqrt{6}x/\sqrt{1+3x^2}$, and $x\in [-1,1]$. Since $\chi_2(x)$ has only one simple zero ($x=0$) on this interval, we can use the following relation: $\delta(\chi_2(x))=\delta(x)/|\chi_2'(0)|$. However, as argued in Sect.~\ref{sec:DD_RBD}, the contribution of this term after integration is zero.
To get $\delta''(\chi_2(x))$, we use the following identity, applying derivatives of Dirac-$\delta$ on a test function $\phi$, 
\begin{equation}
    <\delta^{(n)}(f(x)), \phi> = (-1)^n\frac{d^n}{dy^n}\Bigl(\phi(g(y))g'(y)\Bigr)\Bigr|_{y=0}\;,
\end{equation}
with $f(x)=\chi_2(x)$ in our case, and  $y=f(x)$ and $g$ is the inverse function such as $g(f(x))=x$. For $n=2$, it follows:
\begin{align}
    <\delta''(f(x)), \phi>\, & =\, \frac{d²}{dy^2}\Bigl(\phi(g(y))g'(y)\Bigr)\Bigr|_{y=0}\nonumber\\
    & =\, \left\{g'(y)^3\phi''(g(y))\,+\,3g'(y)g''(y)\phi'(g(y))\right.\nonumber\\
    & \qquad \left.+\,\phi(g(y))g^{(3)}(y)\right\}\Bigr|_{y=0}\nonumber\\
    & =\, \frac{1}{f'(0)^3}\phi''(0) -3\frac{f''(0)}{f'(0)^4}\phi'(0)\nonumber\\
    &\qquad -\,\frac{1}{f'(0)^3}\Bigl[\frac{f^{(3)}(0)}{f'(0)}-3\frac{f''(0)^2}{f'(0)^2}\Bigr]\phi(0)
\end{align}
Using the first and second derivatives of the test function, which in our case is $\phi(x)=x^2/\sqrt{\nu'(x)}=x^2 2\sqrt{3/(1+3x^2)}$ (see Eq.~\eqref{eq:RBD_scaling_int}), we get:
\begin{align}
    \phi'(x) &= 2\sqrt{3}\frac{x(3x^2+2)}{(1+3x^2)^{3/2}}\\
    \phi''(x) &= 2\sqrt{3}\frac{2-3x^2}{(1+3x^2)^{5/2}} \;,
\end{align}
we thus have $\phi(0) = 0$, $\phi'(0) = 0$ and $\phi''(0)=4\sqrt{3}$. This gives
\begin{equation}
    <\delta''(\chi_2(x)), \phi(x)>\, 
    =\, \frac{1}{f'(0)^3}\phi''(0)\, 
    =\, \frac{\sqrt{2}}{3}
\end{equation}
Eventually, Eq.~\eqref{eq:RBD_scaling_int} becomes :
\begin{align}
    K_\perp&= \frac{\sqrt{\pi}}{2}\int \dd k \int_{-1}^1 \dd x \frac{\sqrt{2\pi} \pi k P(k)x^2}{B_0\sqrt{\nu'(x)}} \left(\delta(\chi_2)+\frac{\delta B^2}{2 B_0^2}\delta''(\chi_2)\right)\nonumber\\
    &=\frac{\pi^2}{6} {\delta B^2 \over B_0^3}\int k P(k)dk\nonumber\\ 
    &= \frac{l_c}{24}\frac{\delta B^4}{B_0^3}\,,
\end{align}
from which we get the scaling of the RBD closure in magnetosonic-like polarised turbulence.
\end{document}